%% file: main.tex
\documentclass[english]{article}
\usepackage[OT1]{fontenc}
\usepackage[latin9]{inputenc}
\usepackage{float}
\usepackage{setspace}
\RequirePackage{amsthm}
\RequirePackage[cmex10]{amsmath}
\usepackage{pgf,tikz}
\usetikzlibrary{shapes,decorations,arrows,calc,arrows.meta,fit,positioning}
\tikzset{
	-{Latex[length=5mm, width=1.3mm]},auto,node distance =1 cm and 1 cm,semithick,
	state/.style ={ellipse, draw, minimum width = 0.7 cm},
	point/.style = {circle, draw, inner sep=0.04cm,fill,node contents={}},
	bidirected/.style={dashed,arrows={Latex[length=6mm, width=1.2mm]-Latex[length=4mm, width=1.4mm]}},
	el/.style = {inner sep=2pt, align=left, sloped}
}
\usepackage{amssymb}
\usepackage{ctable}
\usepackage{colortbl}
\usepackage{footnote}
\usepackage{graphicx}
\usepackage[authoryear]{natbib}
\RequirePackage[colorlinks,citecolor=blue,urlcolor=blue]{hyperref}
\PassOptionsToPackage{normalem}{ulem}
\usepackage{ulem}
\usepackage[position=top]{subfig}
\numberwithin{equation}{section}
\theoremstyle{plain}

\numberwithin{equation}{section}

\makeatletter

\newcommand{\indep}{\perp \!\!\! \perp}

\@ifundefined{showcaptionsetup}{}{%
	\PassOptionsToPackage{caption=false}{subfig}}
\usepackage{subfig}
\makeatother
\usepackage{xcolor}

\makeatletter
\newcommand{\globalcolor}[1]{%
	\color{#1}\global\let\default@color\current@color
}
\makeatother

\AtBeginDocument{\globalcolor{black}}

\begin{document}
	
	\title{Controlling for Latent Confounding with Triple Proxies}
	\author{Ben Deaner \thanks{University College London. Email at bendeaner@gmail.com.}}
	\maketitle

\input{abstract}

	\input{introduction2}

		\input{motivatingexample}
	\input{HuSchennachSection}

\input{identification5}

	\input{normalize2}

	\input{conclusion}
	
	\bibliographystyle{authordate1}
	\bibliography{manyproxiesrefs}
	
	\input{partial}
	\input{proofs}
	
\end{document}

%% file: abstract.tex
\begin{abstract}
	We present new results for nonparametric identification of causal effects using noisy proxies for unobserved confounders.  Our approach builds on the results of \citet{Hu2008} who tackle the problem of general measurement error. We call this the `triple proxy' approach because it requires three proxies that are jointly independent conditional on unobservables. We consider three different choices for the third proxy: it may be an outcome, a vector of treatments, or a collection of auxiliary variables. We compare to an alternative identification strategy introduced by  \citet{Miao2018a} in which causal effects are identified using two conditionally independent proxies. We refer to this as the `double proxy' approach. The triple proxy approach identifies objects that are not identified by the double proxy approach, including some that capture the variation in average treatment effects between strata of the unobservables. Moreover, the conditional independence assumptions in the double and triple proxy approaches are non-nested.
	\end{abstract}

%% file: introduction2.tex
A key challenge for causal inference is to credibly adjust for all confounding factors -variables that impact both treatments and outcomes. Unfortunately, important confounders may not be recorded in available data and researchers must make do with noisy and biased proxies for these factors.\footnote{We use the term `proxy' very generally to refer to observables that are statistically associated with some unobservables.} Methods that adjust for observed confounding like inverse propensity score re-weighting
or nonparametric regression typically fail to recover causal effects when some of the confounders are replaced with noisy proxies. This is a problem of potentially non-classical measurement error, in which the mismeasured variables are controls.

For concreteness, suppose we wish to assess the effect of an educational intervention on a student's high school GPA. We adjust for observed pre-treatment characteristics like family background and age of the student. In addition, we would like to control for academic aptitude which could confound treatments and outcomes. While we do not observe aptitude directly, we may have access to test scores which are noisy and possibly biased proxies for academic ability. Simply controlling for test scores and pre-treatment covariates will generally not recover a causal effect.

In this work we provide new results on the nonparametric identification of causal effects using proxies for unobserved confounders. In order to deal with the mismeasured confounding we apply results from \citet{Hu2008} (hereon HS). HS has been applied to numerous problems in empirical economics, notably to the analysis of skill-formation in \citet{Cunha}. HS identify the joint distribution of the observables and underlying mismeasured variables under non-classical measurement error. In order to achieve this, HS require three vectors of proxies that are jointly independent conditional on the latent mismeasured variables. We show that under appropriate conditions either the outcome or vector of treatments can serve as the third proxy. Due to the use of three vectors of proxies we refer to identification based on HS as the `triple proxy' approach.

Crucially, in our setting only confounders are mismeasured. We are interested in effects of treatments, which are measured correctly. This allows us to drop a key assumption of HS, namely that one vector of proxies is a mean- or median-unbiased signal for the unobservables. Without this condition we can only identify distributions involving the latent confounders up to an unknown one-to-one transformation of these factors. However, many causal estimands are invariant to such transformations. For example, the average treatment effect or the effect of treatment on the treated. Therefore, we are able to point identify these objects without an unbiasedness condition. The same is true of some objects that quantify the degree of heterogeneity in treatment effects between different strata of the latent confounders, e.g., the variance of the conditional average treatment effect.  

Depending on the choice of the third proxy (either the outcome, treatments, or a vector of auxiliary variables), identification of causal effects may require a two-step approach. First we identify distributions involving the latent confounders up to an invertible transformation of these factors using results from HS. We then identify objects of interest from a linear integral equation that involves an object obtained in the first step. A closely related two-step strategy was previously explored in the context of regression discontinuity design in  \citet{Rokkanen2015}.

We show that by carefully selecting the third proxy and possibly controlling for treatments or outcomes in the HS step, our approach can accommodate a wide variety of causal relationships between proxies, treatments, and outcomes. This is important because it can be difficult to rule out a priori the possibility that say, a pre-treatment proxy determines treatment, or a post-treatment proxy is affected by treatment. In order to thoroughly and systematically assess the sets of modeling assumptions compatible with our approach, we make extensive use of Directed Acyclic Graphs (DAGs). Our identification results are based on conditional independence restrictions, however the DAGs (or more precisely, the nonparametric structural models associated with each DAG) serve as intuitive  primitive conditions for these independence assumptions.


We compare our results to those of a closely related literature on nonparametric identification using two conditionally independent proxies for latent confounders. We refer to this as the `double proxy' approach. The double proxy approach constitutes a large and recent literature, key papers include  \citet{Miao2018a}, \citet{Deaner2021}, \citet{Kallus2021}, and \citet{Singh2020}. A comparative advantage of the triple proxy approach is that it can identify heterogeneity in causal effects between strata of the latent factors, this is not identified by the double proxy approach. We show that the exclusion restrictions (and consequent conditional independence assumptions) of the double and triple proxy approaches are non-nested. That is, there are conditions under which the double proxy approach is applicable but not the triple proxy approach, and vice versa. Thus our work expands the settings in which one can credibly identify causal effects using proxies for unobserved confounders.

While causal effects of the confounders are not of primary interest in our analysis, an advantage of the triple proxy over the double proxy approach is that under additional restrictions it can be adapted straightforwardly to identify these effects. In a recent work, \citet{Freyberger2021} shows that if the mean- or median-unbiasedness condition of HS is replaced with a related monotonicity condition, then one can identify objects involving quantiles of the latent variables. We apply a similar strategy to \citet{Freyberger2021} in our setting and thus identify the causal effect of shifting the latent confounders between quantiles.

In Section 1 of the paper we provide a motivating example for our analysis. Section 2 contains a brief summary of results in \cite{Hu2008}. Section 3 establishes identification using various choices of the third proxy (treatment, outcomes, or auxiliary variables) and discusses the requisite exclusion restrictions. Section 4 shows how, under additional conditions we can extend our results to identify a richer set of objects including causal effects of the latent variables themselves. Section 5 concludes. The appendix contains proofs as well as additional results that show we can achieve partial identification (and possibly point identification) of causal effects under weaker exclusion restrictions so long as a certain rank invariance condition holds.

\subsection*{Notation and Technicalities}

Throughout we use upper case letters to denote random variables while the corresponding lower case letters denote values of these random variables. If $A$, $B$, $C$, and $D$ are random variables and $A$ and $B$ admit a joint probability density function conditional on $C$ and $D$, then $f_{AB|CD}(a,b|c,d)$ denotes such a density evaluated at $A=a$, $B=b$, $C=c$ and $D=d$.

If $Y$ is an outcome variable and $X$ a treatment, then $Y(x)$ is the potential outcome from a counterfactual level $x$ of $X$. Throughout we implicitly assume that $Y(X)=Y$, a condition sometimes known as `consistency'.

As a technical note, statements about probability densities of random variables should be interpreted to hold for \textbf{some} density compatible with the joint probability measure of these variables. For example, the statement `$f_{AB}(a,b)=f_A(a)f_B(b),\,\forall a,b$' should be understood to mean that `$A$ and $B$ admit a joint density $f_{AB}$ and marginal densities $f_A$ and $f_B$ so that $f_{AB}(a,b)=f_A(a)f_B(b),\forall a,b$'.

%% file: motivatingexample.tex
\section{A Motivating Example}

Suppose we are interested in the causal effect of an educational intervention $X$ on a student's GPA at the end of high-school $Y$. Whether or not a student receives the intervention is determined by the student's teachers,  parents, and perhaps the student herself. These actors base their decision, at least in part, on their private assessments of the student's academic aptitude.

In this setting, academic aptitude (at the time treatment status is decided) is an unobserved confounder. It affects the decision to treat the student and it has an effect on high-school GPA, regardless of treatment. The researcher has access to some test scores that reflect academic ability, but which do not measure it perfectly. Note that our analysis does not require aptitude be one-dimensional, rather it can be understood as a finite-dimensional vector.

In sum, test scores are noisy and possibly biased measurements of an unobserved confounder (academic aptitude). The need to account for the mismeasurement of ability arises in numerous empirical applications, for example in \citet{Griliches1972}, \citet{Fruehwirth2016}, and \cite{Deaner2021}.

Of course, there may be other potential confounders, for example the student's socio-economic characteristics, which are correctly measured in the data. We abstract away from this by implicitly assuming that the researcher has already conditioned on these factors (i.e., our analysis applies within each stratum of these covariates).

Identification is complicated by the fact that test scores can directly cause or be caused by treatments and outcomes. If the educational intervention affects a student's academic progress, then it presumably affects the scores on tests taken after the intervention. If a test is taken prior to the intervention, then it may determine eligibility for treatment. A test score could directly enter into the GPA calculation or may be used to decide some feature of the student's education other than the intervention, and thus it may have an affect on GPA that is not mediated by treatment.

Ruling out causal relationships of this kind requires detailed institutional knowledge. In this work we show that the triple proxy approach allows for causal relationships between the proxies, treatments, and outcomes that are incompatible with the double proxy approach, and vice versa. Thus there may be settings in which institutional knowledge is compatible with one of the two approaches but not the other.

In Figure 1 we present Directed Acyclic graphs (DAGs) that encode possible causal relationships between academic ability, the treatment (an educational intervention), the outcome (GPA at the end of high-school), and two or three sets of test scores.

Each  DAG encodes
a set of exclusion restrictions in a Non-Parametric Structural Equations Model (NPSEM) of the kind in \citet{Pearl2009}. Each NPSEM  implies a set of conditional independence restrictions required for identification of causal effects using either the double or triple proxy approach.

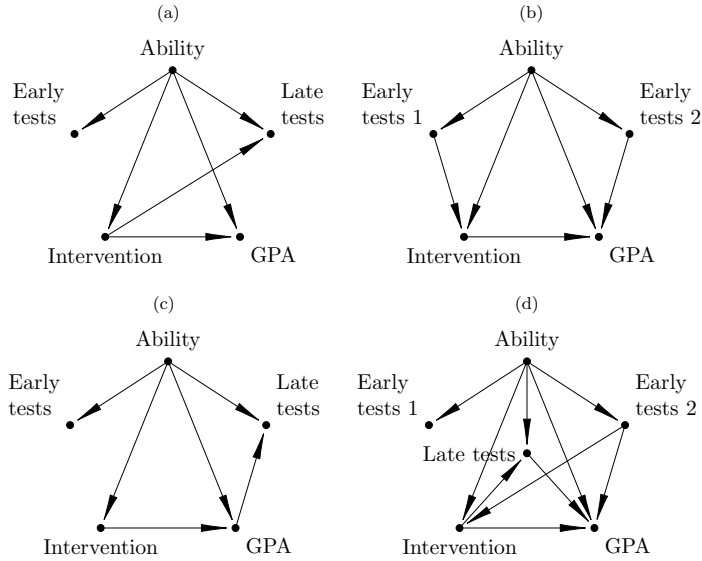
\begin{figure}[h]
\centering
	\caption{Test Score Proxy Graphs}

	\resizebox{!}{11pt}{%
		\subfloat[]{
						\begin{tikzpicture}
			\node (y) at (1.1,-1.22) [label={[align=left]below right:GPA},point];
	\node (x) at (-1.1,-1.22) [label={[align=left]below :Intervention},point];
	\node (w) at (0,1.5) [label={[align=left]above:Ability},point];
	\node (v) at (1.6,0.46) [label={[align=left]above right:Late\\tests},point];
	\node (z) at (-1.6,0.46) [label={[align=left]above left:Early\\tests},point];
	
	\path (x) edge (y);
	\path (w) edge (x);
	\path (w) edge (y);
	\path (w) edge (z);
	\path (w) edge (v);
	\path (x) edge (v);
				\end{tikzpicture}
	}	
}
	\resizebox{!}{11pt}{%
		\subfloat[]{
			\begin{tikzpicture}
				
			\node (y) at (1.1,-1.22) [label={[align=left]below right:GPA},point];
\node (x) at (-1.1,-1.22) [label={[align=left]below :Intervention},point];
\node (w) at (0,1.5) [label={[align=left]above:Ability},point];
\node (v) at (1.6,0.46) [label={[align=left]above right:Early\\tests 2},point];
\node (z) at (-1.6,0.46) [label={[align=left]above left:Early\\tests 1},point];
				
		\path (x) edge (y);
\path (w) edge (x);
\path (w) edge (y);
\path (w) edge (z);
\path (w) edge (v);
\path (z) edge (x);
\path (v) edge (y);
			\end{tikzpicture}
		}	
}

\resizebox{!}{11pt}{%
			\subfloat[]{		
		\begin{tikzpicture}
			
			\node (y) at (1.1,-1.22) [label={[align=left]below right:GPA},point];
			\node (x) at (-1.1,-1.22) [label={[align=left]below :Intervention},point];
			\node (w) at (0,1.5) [label={[align=left]above:Ability},point];
			\node (v) at (1.6,0.46) [label={[align=left]above right:Late\\tests},point];
			\node (z) at (-1.6,0.46) [label={[align=left]above left:Early\\tests},point];

			\path (x) edge (y);
			\path (w) edge (x);
			\path (w) edge (y);
			\path (w) edge (z);
			\path (w) edge (v);
			\path (y) edge (v);
		\end{tikzpicture}
	}
}
\resizebox{!}{11pt}{%
		\subfloat[]{	
	\begin{tikzpicture}
	
			\node (y) at (1.1,-1.22) [label={[align=left]below right:GPA},point];
\node (x) at (-1.1,-1.22) [label={[align=left]below :Intervention},point];
\node (w) at (0,1.5) [label={[align=left]above:Ability},point];
\node (v) at (1.6,0.46) [label={[align=left]above right:Early\\tests 2},point];
\node (z) at (-1.6,0.46) [label={[align=left]above left:Early\\tests 1},point];
\node (c) at (0,0) [label={[align=left]left:Late tests },point];
	
	\path (x) edge (y);
	\path (w) edge (x);
	\path (w) edge (y);
	\path (w) edge (z);
	\path (w) edge (v);
	\path (w) edge (c);
	\path (v) edge (y);
	\path (c) edge (y);
	\path (v) edge (x);
	\path (x) edge (c);
\end{tikzpicture}
}
}
\end{figure}

The causal diagram in Figure 1.a implies the conditional independence restrictions required by both the double and triple proxy approaches. In this DAG one set of scores are from `early tests' taken prior to the decision to treat and some are from `late tests' taken after treatment is administered.

The DAG in Figure 1.a encodes an assumption that there is no direct effect of the tests on high-school GPA, nor on treatment. This effectively rules out the possibility that the test scores are used to determine any important aspects of a student's education. In \citet{Deaner2021} this is justified by the fact that the test scores are only observed by researchers who have no input into the students' education.

The graph in Figure 1.a allows the educational intervention to affect the scores on post-treatment tests, this is important because if the educational intervention affects academic performance then it likely affects future test scores.

The graph in Figure 1.b is adapted from \citet{Miao2018a}.  In this case one set of early tests can determine treatment. The other early tests cannot affect treatment but could impact some other aspect of the student's education and thus affect the outcome.

Figure 1.b is compatible with the double proxy approach but not the triple proxy approach. The triple proxy approach employs \citet{Hu2008}, which requires three proxies that are independent conditional on unobservables. Under Figure 1.b no three of the four observables are guaranteed to be jointly independent conditional on ability. Nor are there three observables that are independent conditional on ability and whichever observable is left over.

Figure 1.c is compatible with the triple proxy approach but not the double proxy approach. In this case, one set of scores are from tests taken after high-school graduation. High-school GPA could affect say, college attendance and thus later test scores. The treatment may affect these late test scores so long as this is mediated by academic progress in high-school as measured by GPA.

The model in Figure 1.d allows for identification using the triple proxy approach but not double proxies. This is because two of the three available proxies have direct causal links to both treatments and outcomes, which rules out their use in the double proxy approach. In this model one set of early tests can determine treatment and may have some other impact on the student's education and thus affect GPA directly. Treatment can impact the late test scores, and these scores may determine the course of a student's later education and thus affect GPA.

%% file: HuSchennachSection.tex
\section{Identification in Hu and Schennach (2008)}

HS prove identification of a nonparametric factor
model. We restate their results below. Let $W$ be an unobserved,
possibly vector-valued latent factor with support $\mathcal{W}$.
Let $V$, $Z$, and $C$ be observable random vectors with respective
supports $\mathcal{V}$, $\mathcal{Z}$, and $\mathcal{C}$. The following
conditions are from Hu and Schennach (2008).

\theoremstyle{definition}
\newtheorem*{AHS1}{HS Assumption 1}
\begin{AHS1}
 $V$, $Z$, $W$, and $C$ admit a bounded, non-zero density with respect to the product measure of the Lebesgue measure on $\mathcal{V}\times\mathcal{Z}\times\mathcal{W}$
and some dominating measure $\mu$ on $\mathcal{C}$. All marginal and conditional densities are also bounded.
\end{AHS1}

\theoremstyle{definition}
\newtheorem*{AHS2}{HS Assumption 2}
\begin{AHS2}
$V$, $Z$, and $C$ are jointly independent conditional on $W$. Formally, $V\indep(Z,C)|W$ and $Z\indep C|W$.
\end{AHS2}

\theoremstyle{definition}
\newtheorem*{AHS3}{HS Assumption 3}
\begin{AHS3}
For any bounded function $\delta$ in $ L_1(\mathcal{W})$:
\[\int_\mathcal{W} f_{V|W}(V|w)\delta(w)dw\overset{a.s.}{=}0\implies \delta(W)\overset{a.s.}{=}0\]
and the same holds with $V$ replaced by $Z$.
\end{AHS3}

\theoremstyle{definition}
\newtheorem*{AHS4}{HS Assumption 4}
\begin{AHS4}
For any $w_1,w_2\in\mathcal{W}$ if $w_1\neq w_2$ then $P\big(f_{C|W}(C|w_1)\neq f_{C|W}(C|w_2)\big)>0$.
\end{AHS4}

HS Assumption 1 ensures some bounded densities exist. HS Assumption 2 states that $V$, $Z$, and $C$ are jointly independent conditional on $W$.

If the marginal densities $f_W$ , $f_V$, and $f_Z$ are  bounded below away from zero over $\mathcal{W}$, $\mathcal{V}$, and $\mathcal{Z}$ respectively, then Assumption 3 is equivalent to two bounded completeness conditions. Namely, bounded completeness of $W$ for $V$, and bounded completeness of $W$ for $Z$. Note that Assumption 3 differs slightly from the corresponding condition in \citet{Hu2008}, the version we use here is employed in the Handbook of Econometrics treatment of HS (see \citet{Schennach2020}).

Statistical completeness conditions are used to identify Nonparametric Instrumental Variables (NPIV) models of the kind in \citet{Newey2003} and \citet{Ai2003}. Thus condition 3 states that $V$ and $Z$ are both relevant instruments for $W$ in the sense of NPIV.

Assumptions 4 is a relatively weak condition on the association between $C$ and $W$. \citet{Hu2008} note that this assumption is weaker than imposing HS Assumption 3 on $C$. In words it states that any change in $W$ must induce some change in the conditional distribution of $C$. This condition can hold even if $C$ is a binary random variable and $W$ is a continuous random vector.

\theoremstyle{plain}
\newtheorem*{HST}{HS Theorem (Hu and Schennach (2008))}
\begin{HST}
Under HS Assumptions 1 and 2 the following equality holds:
\begin{equation}
f_{ZC|V}(z,c|v)=\int_\mathcal{W}\label{id1} f_{C|W}(c|w)f_{W|V}(w|v)f_{Z|W}(z|w)dw
\end{equation}
Moreover, under HS Assumptions 1-4, $f_{W|V}$, $f_{Z|W}$,
and $f_{W|C}$ are identified from the above up to a reordering of $W$. 
\end{HST}

To formalize what we mean by `identified up to reorderings', suppose some other conditional densities $\tilde{f}_{W|V}$, $\tilde{f}_{Z|W}$,
and $\tilde{f}_{W|C}$ satisfy Assumption 1-4 and (\ref{id1}):
\begin{align*}
	f_{ZC|V}(z,c|v)&=\int_\mathcal{W} \tilde{f}_{W|V}(w|v)\tilde{f}_{Z|W}(z|w)\tilde{f}_{C|W}(c|w)dw\end{align*}

Then there exists an bijective function $\varphi:\mathcal{W}\to\mathcal{W}$ so that $\tilde{f}_{W|V}(w|v)=f_{\varphi(W)|V}(w|v)$, $\tilde{f}_{Z|W}(z|w)=f_{Z|\varphi(W)}(z|w)$, and $\tilde{f}_{C|W}(c|w)=f_{C|\varphi(W)}(c|w)$.

Theorem 1 identifies conditional densities up to a reordering of $W$. HS pin down a single ordering using an additional assumption given below. In the case of mismeasured control variables, causal effects of interest are often invariant to reordering of $W$, and so we do not require this condition. However, we revisit it in Section 4.

\theoremstyle{definition}
\newtheorem*{AHS5}{HS Assumption 5}
\begin{AHS5}
	There is a known functional $M$ so that $M[f_{Z|W}(\cdot|w)]=w,\,\forall w\in\mathcal{W}$.
\end{AHS5}

If the functional $M$ returns the mean of the distribution in its argument then the assumption states that $Z$ is mean-unbiased for $W$. If $M$ returns the median, then the assumption states that $Z$ is median-unbiased for $W$. It is implicit in the assumption that the dimensions of $W$ and $Z$ are the same.

%% file: identification5.tex
\section{Identifying Causal Effects}

To identify causal effects with mismeasured controls, we suppose that two vectors of proxies $V$ and $Z$ are available. The third proxy $C$ will be either the outcome $Y$, treatment $X$, or some additional observables. These choices are appropriate under different sets of exclusion restrictions.

\subsubsection*{Objects of Interest}

As discussed in the previous section, without a condition like HS Assumption 5, we can only identify objects involving $W$ up to a reordering of this variable. Fortunately, causal effects of the treatment $X$ are typically invariant to  reordering of $W$, and so we can point identify many causal objects without invoking such an assumption. These include objects that capture the degree of heterogeneity in treatment response among groups with different values of $W$. Such objects are not identified using the double proxy approach.

In order to point identify causal objects, we first identify the joint distributions of potential outcomes $Y(x)$, latent variables $W$, and realized treatments $X$, for $\mu_X$-almost all $x$, up to a reordering of $W$.\footnote{$\mu_X$ and $\mu_Y$ are a probability laws that dominates those of $X$ and $Y$ respectively. We allow for both discrete and continuous treatments and likewise for the outcome.}

Let $f_{Y(x)WX}$ be a joint density of $Y(x)$, $W$, and $X$. The marginal density of potential outcomes and the density conditional on the realized treatment are given by the expressions below. These are invariant to reordering of $W$ and so can be point identified even when we cannot recover the ordering of $W$.
\begin{align}
	f_{Y(x_{1})|X}(y|x_{2})&=\int_{\mathcal{W}}f_{Y(x_1)W|X}(y,w|x_{2})dw\label{yx1}\\
	f_{Y(x_1)}(y) &=\int_\mathcal{X} f_{Y(x_{1})}(y|x)f_X(x)d\mu_X(x)\label{yx2}
\end{align}

From the above we can further obtain say, average treatment effects and the effect of treatment on the treated, as well as  quantile treatment effects.\footnote{By `quantile treatment effects' we mean the differences in quantiles of potential outcomes under different treatments. These may differ from the quantiles of the treatment effect.} With continuous treatments we can identify average partial effects under a suitable differentiability condition.
 
In addition, from $f_{Y(x)WX}$ we can recover objects that capture the heterogeneity in treatment response between groups with different values of the latent variables $W$.  Suppose $X$ is binary, then the conditional average treatment effect $\beta(W)\equiv E[Y(1)-Y(0)|W]$, is given by:
\[
\beta(w)=\int_\mathcal{Y} y f_{Y(1)|W}(y|w)d\mu_Y-\int_{\mathcal{Y}} y f_{Y(0)|W}(y|w)d\mu_Y
\]
Where we implicitly assume the mean exists and is finite. While $\beta(w)$ itself is not invariant to reorderings of $W$, its distribution is invariant. Thus we can identify say, the variance of $\beta(W)$. The cumulative distribution function (CDF) of $\beta(W)$ conditional on $X=x$, and the marginal CDF are given below:
\begin{align}
F_{\beta(W)|X}(b|x)&=\int_{\mathcal{W}} 1\{\beta(W)\leq a\} f_{W|X}(w|x)dw\label{tau1}\\
F_{\beta(W)}(b)&=\int_{\mathcal{X}} F_{\beta(W)|X}(b|x) d\mu_X(x) \label{tau2}
\end{align}

Neither the marginal nor conditional distribution of $\beta(W)$ are identified by the double proxy approach.

Similarly, when treatment is continuous, and the derivatives of potential outcomes exist and are bounded, we can obtain the conditional average partial effect $\pi(W)\equiv E[\frac{\partial}{\partial x} Y(x)|W]$ up to a reordering of $W$ as follows:
\[ 
\pi(W)=\frac{\partial}{\partial x}\int_{\mathcal{Y}} y F_{Y(x)|W}(dy|W)
\]
The conditional and marginal CDFs of the conditional average partial effect are then given by:
\begin{align*}
F_{\pi(W)|X}(p|x)&=\int_{\mathcal{W}} 1\{\pi(W)\leq p\} f_{W|X}(w|x)dw\\
F_{\pi(W)}(p)&=\int_{\mathcal{X}} F_{\pi(W)|X}(p|x)d\mu_X(x)
\end{align*}

As a technical caveat, when treatments are continuous we can only identify the densities $f_{Y(x)WX}$ for $\mu_X$-almost all $x$ (as opposed to all $x$), up to a reordering of $W$. In practice this is likely to be of little consequence because for estimation one typically assumes the support of $X$ is rectangular and objects of interest like $E[Y(x)]$ are continuous in $x$. In this case, if $E[Y(x)]$ is identified almost everywhere then it is in fact identified everywhere.

\subsection{Outcome Proxies}

We first apply the results in Section 2 to identify $f_{Y|WX}$ and $f_{WX}$ up to a reordering of $W$ in a first stage. The outcome $Y$ acts as the proxies $C$. We apply the results in Section 2 after conditioning on $X$, i.e., within each stratum of the treatment. Having identified $f_{Y|WX}$ and $f_{WX}$ we recover $f_{Y(x)WX}$ up to a reordering. We can then point identify causal objects like (\ref{yx2}) which are invariant to reordering of $W$.

In this context we assume the existence of bounded densities akin to HS Assumption 1 in the previous section. We also assume that at each $x$, the potential outcome function $Y(\cdot)$ is almost surely continuous. This assumption is trivially true when treatment is discrete and allows us to avoid some technical issues that arise with continuous treatments. Let $\mathcal{V}$, $\mathcal{Z}$, $\mathcal{W}$, $\mathcal{Y}$, and $\mathcal{X}$ be the supports of $V$, $Z$, $W$, $Y$, and $X$ respectively.

\theoremstyle{definition}
\newtheorem*{A1}{Assumption 1}
\begin{A1}
	$V$, $Z$, $W$, $Y$, and $X$ admit a bounded, non-zero density with respect to
	the product of	the Lebesgue measure on $\mathcal{V}\times\mathcal{Z}\times\mathcal{W}$, some dominating measure $\mu_Y$ on $\mathcal{Y}$,
	and a dominating measure $\mu_X$ on $\mathcal{X}$.
	All marginal and conditional densities are also bounded.
\end{A1}

Figure 2 contains three alternative causal graphs. These graphs encode exclusion restrictions on an NPSEM that imply a set of conditional independence restrictions which we use for identification. 

\begin{figure}[h]
	\centering
	\caption{Outcome Proxy Graph}
	
	\centering
				\resizebox{!}{11pt}{%
		\subfloat[]{
					
	\begin{tikzpicture}
		
		\node (y) at (1.1,-1.22) [label=below right:$Y$,point];
		\node (x) at (-1.1,-1.22) [label=below left:$X$,point];
		\node (w) at (0,1.5) [label=above:$W$,point];
		\node (v) at (1.6,0.46) [label=above right:$V$,point];
		\node (z) at (-1.6,0.46) [label=above left:$Z$,point];
		
		\path (x) edge (y);
		\path (w) edge (x);
		\path (w) edge (y);
		\path (w) edge (z);
		\path (w) edge (v);
		\path (v) edge (x);
		
	\end{tikzpicture}
}
		\subfloat[]{
	\begin{tikzpicture}
	
	\node (y) at (1.1,-1.22) [label=below right:$Y$,point];
	\node (x) at (-1.1,-1.22) [label=below left:$X$,point];
	\node (w) at (0,1.5) [label=above:$W$,point];
	\node (v) at (1.6,0.46) [label=above right:$V$,point];
	\node (z) at (-1.6,0.46) [label=above left:$Z$,point];
	
	\path (x) edge (y);
	\path (w) edge (x);
	\path (w) edge (y);
	\path (w) edge (z);
	\path (v) edge (w);
	\path (v) edge (x);
	
\end{tikzpicture}
}
		\subfloat[]{
	
	\begin{tikzpicture}
		
		\node (y) at (1.1,-1.22) [label=below right:$Y$,point];
		\node (x) at (-1.1,-1.22) [label=below left:$X$,point];
		\node (w) at (0,1.5) [label=above:$W$,point];
		\node (v) at (1.6,0.46) [label=above right:$V$,point];
		\node (z) at (-1.6,0.46) [label=above left:$Z$,point];
		
		\path (x) edge (y);
		\path (w) edge (x);
		\path (w) edge (y);
		\path (w) edge (z);
		\path (w) edge (v);
		\path (x) edge (v);
		
	\end{tikzpicture}
}
}	
\end{figure}
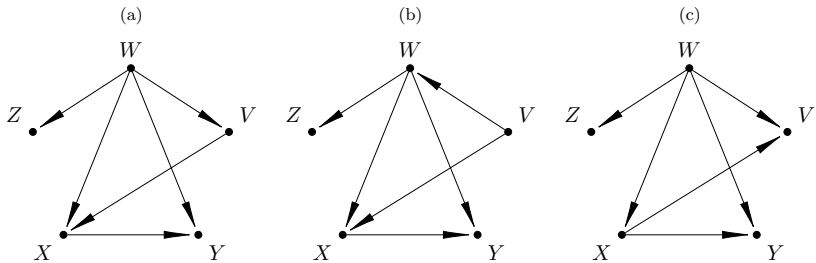	
	
The causal graphs in Figures 2.a and 2.b suggest that $V$ is a pre-treatment variable and allow $V$ to affect the treatment $X$. The graph in 2.c suggests $V$ is a post-treatment and could be affected by treatment.

The graphs preclude $X$ affecting $Z$ or vice versa. This is most credible when $Z$ is a pre-treatment variable (and thus cannot be affected by treatment), which is not used to decide treatment.

Crucially, neither the proxies $Z$ nor $V$ may directly affect the outcome $Y$. Moreover, $Z$ must not directly affect $V$ nor vice versa.

The graphs in Figure 2 imply a set of conditional independence restrictions given in Proposition 1. The proposition can be verified straight-forwardly using the tools in \citet{Pearl2009}. Our identification results directly assume the conditional independence restrictions in the conclusion of Proposition 1. Thus the graphs in Figure 2 can be understood to represent primitive conditions for these conditional independence restrictions.

\theoremstyle{plain}
\newtheorem*{P1}{Proposition 1}
\begin{P1}
The NPSEMs associated with the causal graphs in Figure 2 all imply the following conditional independence restrictions:

i. $Y\indep(V,Z)|(W,X)$, ii. $V\indep Z|(W,X)$, iii. $Z\indep X|W$, and iv. $Y(x)\indep (X,V)|W$
\end{P1}

The conditional independence restrictions in Proposition 1 are stronger than those required for the double proxy approach. In particular, the double proxy approach requires conditions ii., iii., and iv. but not condition i.

In this setting we use the Assumption 2 below in place of HS Assumption 3.

\theoremstyle{definition}
\newtheorem*{A2}{Assumption 2}
\begin{A2}
	For each $x\in\mathcal{X}$ and any bounded function $\delta$ in $ L_1(\mathcal{W})$:
	\[\int_\mathcal{W} f_{V|WX}(V|w,x)\delta(w)dw\overset{a.s.}{=}0,\,\implies \delta(W)\overset{a.s.}{=}0\]
and the same holds with $V$  replaced by $Z$.
\end{A2}

Assumption 2 differs from HS Assumption 3 in that it must hold within each stratum of the treatment $X$. If the conditional densities $f_{W|X}$, $f_{V|X}$, and $f_{Z|X}$ are all bounded below away from zero and have bounded supports, then Assumption 2 is equivalent to the completeness conditions in \citet{Deaner2021}.

Finally, Assumption 3 below plays the role of HS Assumption 4. Note that this condition allows for the possibility that the outcome $Y$ is binary, even if $W$ is a continuous random vector.
\theoremstyle{definition}
\newtheorem*{A3}{Assumption 3}
\begin{A3}
	For all $x\in\mathcal{X}$ and any $w_1,w_2\in\mathcal{W}$, if $w_1\neq w_2$ then:
	\[P\big(f_{Y|WX}(Y|w_1,x)\neq f_{Y|WX}(Y|w_2,x)\big)>0\]
\end{A3}

We now identify causal effects under the conditions above.

\theoremstyle{plain}
\newtheorem*{T1}{Theorem 3.1 (Outcome Proxies)}
\begin{T1}
	Suppose Assumptions 1-3 and conclusions i., ii., and iii. of Proposition 1 and  hold. Then $f_{Y|WX}$, $f_{Z|W}$, and $f_{W|VX}$ (and thus $f_{WX}$) are identified up to a reordering of $W$ from the equation below:
\begin{equation*}
f_{YZ|VX}(y,z|v,x)=\int_\mathcal{W} f_{Y|WX}(y|w,x)f_{W|VX}(w|v,x)f_{Z|W}(z|w)dw
\end{equation*}
Under conclusion iv. of Proposition 1, for $\mu_X$-almost all $x_1$: \begin{equation}f_{Y(x_1)WX}(y,w,x_2)=f_{Y|WX}(y|w,x_1)f_{WX}(w,x_2)\label{eqpot}
\end{equation} 
And so $f_{Y(x)WX}$ is identified for $\mu_X$-almost all $x$, up to a reordering of $W$. Causal objects can then be point identified by say, (\ref{yx1}) or (\ref{yx2}).
\end{T1}

Without Conclusion iii. of Proposition 1, we could still apply \citet{Hu2008} to identify $f_{Y|WX}(\cdot|\cdot,x)$, $f_{Z|WX}(\cdot|\cdot,x)$, and $f_{W|VX}(\cdot|\cdot,x)$ up to a reordering of $W$ for each $x$ in the support of $X$. However, the reordering of $W$ could differ between the values of $x$. We revisit this possibility in Appendix A and show that partial identification (and possibly point identification) can be achieved without condition iii. under a rank invariance assumption.

\subsection{Treatment Proxies}

We now consider the case in which the third proxy $C$, is the vector of treatments $X$. In this case identification proceeds in two stages. In a first stage we use results from HS to identify conditional distributions involving $W$ up to reordering. In a second step, the conditional distribution of potential outcomes is identified via a linear integral equation. This two-step approach is similar to one employed in \citet{Rokkanen2015}.

In this case, the causal diagrams below are sufficient for the conditional independence restrictions under which we establish identification.

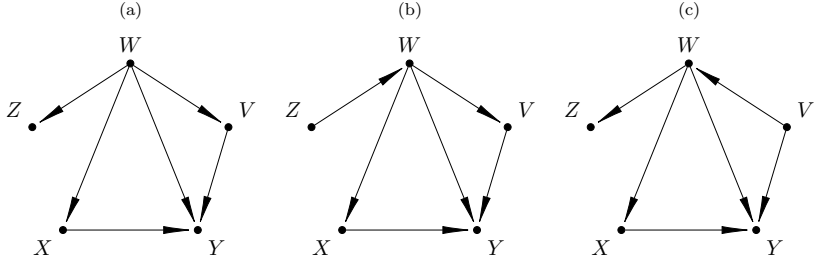
\begin{figure}[h]
	\centering
	\caption{Treatment Proxy Graphs}
	
	\centering
			\resizebox{!}{11pt}{%
	\subfloat[]{		
	\begin{tikzpicture}
		
		\node (y) at (1.1,-1.22) [label=below right:$Y$,point];
		\node (x) at (-1.1,-1.22) [label=below left:$X$,point];
		\node (w) at (0,1.5) [label=above:$W$,point];
		\node (v) at (1.6,0.46) [label=above right:$V$,point];
		\node (z) at (-1.6,0.46) [label=above left:$Z$,point];

 		\path (x) edge (y);
		\path (w) edge (x);
		\path (w) edge (y);
		\path (w) edge (z);
		\path (w) edge (v);
		\path (v) edge (y);
	\end{tikzpicture}
}
	\subfloat[]{
	\begin{tikzpicture}
	
	\node (y) at (1.1,-1.22) [label=below right:$Y$,point];
	\node (x) at (-1.1,-1.22) [label=below left:$X$,point];
	\node (w) at (0,1.5) [label=above:$W$,point];
	\node (v) at (1.6,0.46) [label=above right:$V$,point];
	\node (z) at (-1.6,0.46) [label=above left:$Z$,point];

	\path (x) edge (y);
	\path (w) edge (x);
	\path (w) edge (y);
	\path (z) edge (w);
	\path (w) edge (v);
	\path (v) edge (y);
\end{tikzpicture}
}
	\subfloat[]{
	\begin{tikzpicture}
		
		\node (y) at (1.1,-1.22) [label=below right:$Y$,point];
		\node (x) at (-1.1,-1.22) [label=below left:$X$,point];
		\node (w) at (0,1.5) [label=above:$W$,point];
		\node (v) at (1.6,0.46) [label=above right:$V$,point];
		\node (z) at (-1.6,0.46) [label=above left:$Z$,point];

		\path (x) edge (y);
		\path (w) edge (x);
		\path (w) edge (y);
		\path (w) edge (z);
		\path (v) edge (w);
		\path (v) edge (y);
	\end{tikzpicture}
}
}
	
\end{figure}

The diagrams in Figure 3 suggest that $V$ is determined prior to the outcome, and the graphs allow $V$ to directly affect the outcome. However, $V$ and $Z$ cannot directly affect, or be directly affected by, the treatment $X$. This contrasts with the double proxy case, which allows one of the two proxies to be directly causally connected to the treatment.

\newtheorem*{P2}{Proposition 2}
\begin{P2}
	The NPSEMs associated with the causal graphs in Figure 3 imply the following conditional independence restrictions:

i. $V\indep(X,Z)|W$, ii. $X\indep Z|W$, iii. $Y\indep Z|(W,X)$, and iv. $Y(x)\indep (X,Z)|W$
\end{P2}

Conditions i., and iv. in Proposition 2 are those required for the double proxy approach. However, the double proxy approach does not require condition ii.

Strictly speaking, condition iii. is not required for the double proxy approach. However, if condition iv. is strengthened slightly to $Y(\cdot)\indep (Z,X)|W$ this implies condition iii.

Assumption 4 replaces HS Assumption 4. Note that Assumption 4 may hold even if the treatment $X$ is binary.

\theoremstyle{definition}
\newtheorem*{A4}{Assumption 4}
\begin{A4}
	For any $w_1,w_2\in\mathcal{W}$ if $w_1\neq w_2$ then $P\big(f_{X|W}(X|w_1)\neq f_{X|W}(X|w_2)\big)>0$.
\end{A4}

\theoremstyle{plain}
\newtheorem*{T2}{Theorem 3.2 (Treatment Proxies)}
\begin{T2}
	Suppose conclusions i. and ii. of Proposition 2, HS Assumption 3, and Assumptions 1 and 4 hold. Then $f_{Z|W}$ is identified up to a reordering of $W$ from the equation below:
\begin{equation*}
f_{ZX|V}(z,x|v)=\int_{\mathcal{W}}f_{X|W}(x|w)f_{W|V}(w|v)f_{Z|W}(z|w)dw
\end{equation*}
If conclusion iii. of Proposition 2 also holds, $f_{YWX}$ is then identified up to reorderings of $W$ from the integral equation below.
\begin{equation*}
	f_{YZX}(y,z,x)=\int_{\mathcal{W}}f_{YWX}(y,w,x)f_{Z|W}(z|w)dw
\end{equation*}
Under conclusion iv. of Proposition 2 we recover $f_{Y(x)WX}$ for $\mu_X$-almost all $x$, up to a reordering of $W$ from (\ref{eqpot}).  Causal objects can then be point identified by say, (\ref{yx1}) or (\ref{yx2}).
\end{T2}

\subsection{Outcome-Conditional Treatment Proxies}

We again consider the case in which the third proxy $C$, is the vector of treatments $X$. However, we apply \citet{Hu2008} within each stratum of the outcome. This allows for the possibility that the outcome directly affects one of the proxies $V$, which is generally incompatible with the double proxy approach.

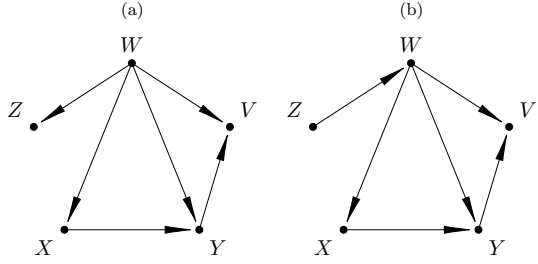
\begin{figure}[h]
	\centering
	\caption{Treatment Proxy Graphs}
	
	\centering
	\resizebox{!}{11pt}{%
		\subfloat[]{		
			\begin{tikzpicture}
				
				\node (y) at (1.1,-1.22) [label=below right:$Y$,point];
				\node (x) at (-1.1,-1.22) [label=below left:$X$,point];
				\node (w) at (0,1.5) [label=above:$W$,point];
				\node (v) at (1.6,0.46) [label=above right:$V$,point];
				\node (z) at (-1.6,0.46) [label=above left:$Z$,point];

				\path (x) edge (y);
				\path (w) edge (x);
				\path (w) edge (y);
				\path (w) edge (z);
				\path (w) edge (v);
				\path (y) edge (v);
			\end{tikzpicture}
		}
		\subfloat[]{
			\begin{tikzpicture}
				
				\node (y) at (1.1,-1.22) [label=below right:$Y$,point];
				\node (x) at (-1.1,-1.22) [label=below left:$X$,point];
				\node (w) at (0,1.5) [label=above:$W$,point];
				\node (v) at (1.6,0.46) [label=above right:$V$,point];
				\node (z) at (-1.6,0.46) [label=above left:$Z$,point];

				\path (x) edge (y);
				\path (w) edge (x);
				\path (w) edge (y);
				\path (z) edge (w);
				\path (w) edge (v);
				\path (y) edge (v);
			\end{tikzpicture}
		}
	}
	
\end{figure}

The diagrams in Figure 4 differ from those in Figure 3 in that $V$ is a post-outcome variable and can be impacted directly by the outcome. Note that $V$ is a post treatment variable but $X$ must not affect $V$ directly. 

Recall the test score example with the outcome $Y$ measuring GPA in the final year of high-school. Suppose the tests in $V$ taken a year after high-school graduation. GPA may affect college attendance which could in turn impact scores on the college-age tests $V$. The educational intervention $X$ can influence post-high school test scores, so long as the effect of the intervention is mediated by academic achievement over high school as measured by final GPA.

\newtheorem*{P3}{Proposition 3}
\begin{P3}
	The NPSEMs associated with the causal graphs in Figure 4 imply the following conditional independence restrictions:
	
	i. $V\indep(X,Z)|(W,Y)$, ii. $X\indep Z|(W,Y)$, iii. $Y\indep Z|W$, and iv., $Y(x)\indep X|W$.
\end{P3}

The conditions in Proposition 3 are insufficient for the double proxy approach. The double proxy approach requires that $V\indep(X,Z)|W$ which generally rules out $V$ having a direct causal effect on $Y$ (unless we were to assume $X$ has no causal effect on $Y$ which defeats the purpose of our analysis). Conversely, the double proxy approach does not require any independence between $X$ and $Z$, conditional or otherwise.

In this setting we need HS Assumption 3 to apply within each stratum of the outcome.

\theoremstyle{definition}
\newtheorem*{A5}{Assumption 5}
\begin{A5}
	For each $y\in\mathcal{Y}$ and any bounded function $\delta$ in $ L_1(\mathcal{W})$:
\[\int_\mathcal{W} f_{V|WY}(V|w,y)\delta(w)dw\overset{a.s.}{=}0,\,\implies \delta(W)\overset{a.s.}{=}0\]
and the same holds with $V$  replaced by $Z$.
\end{A5}

Finally, we need HS Assumption 4 to hold for the treatment proxy within each stratum of $Y$.

\theoremstyle{definition}
\newtheorem*{A6}{Assumption 6}
\begin{A6}
	For all $y\in\mathcal{Y}$ and any $w_1,w_2\in\mathcal{W}$, if $w_1\neq w_2$ then:
\[P\big(f_{X|WY}(X|w_1,y)\neq f_{X|WY}(X|w_2,y)\big)>0\]
\end{A6}

\theoremstyle{plain}
\newtheorem*{T3}{Theorem 3.3 (Conditional Treatment Proxies)}
\begin{T3}
	
Suppose conclusion i., ii., and iii, of Proposition 3 and Assumptions 1, 5, and 6 hold. Then $f_{X|WY}$, $f_{Z|W}$, and $f_{W|VY}$ are identified up to a reordering of $W$ from the equation below:
\begin{equation*}
	f_{XZ|VY}(x,z|v,y)=\int_{\mathcal{W}}f_{X|WY}(x|w,y)f_{W|VY}(w|v,y)f_{Z|WY}(z|w,y)dw
\end{equation*}
$f_{YWX}$ can be written in terms of $f_{X|WY}$, $f_{W|VY}$, and $f_{VY}$. If conclusion iv. of Proposition 3 we recover $f_{Y(x)WX}$ for $\mu_X$-almost all $x$, up to a reordering of $W$ from (\ref{eqpot}). Causal objects can then be point identified by say, (\ref{yx1}) or (\ref{yx2}).	
	
\end{T3}

\subsection{Auxiliary Proxies}

Finally, we consider the case in which the third proxy $C$ is a vector of auxiliary variable (as opposed to $X$ or $Y$). In this case we apply the results from Section 2 within each stratum of the treatments.

\theoremstyle{definition}
\newtheorem*{A7}{Assumption 7}
\begin{A7}
	$V$, $Z$, $W$, $Y$, $X$, and $C$ admit a bounded, non-zero density with respect to
	the product of	the Lebesgue measure on $\mathcal{V}\times\mathcal{Z}\times\mathcal{W}$, some dominating measure $\mu_Y$ on $\mathcal{Y}$,
	 a dominating measure $\mu_X$ on $\mathcal{X}$, and a dominating measure $\mu_C$ on $\mathcal{C}$.
	All marginal and conditional densities are also bounded.
\end{A7}

\begin{figure}[h]
	\centering
	\caption{Auxiliary Proxy Graphs}
	
	\centering	
				\resizebox{!}{11pt}{%
\subfloat[]{
			\begin{tikzpicture}
				
				\node (y) at (1.1,-1.22) [label=below right:$Y$,point];
				\node (x) at (-1.1,-1.22) [label=below left:$X$,point];
				\node (w) at (0,1.5) [label=above:$W$,point];
				\node (v) at (1.6,0.46) [label=above right:$V$,point];
				\node (z) at (-1.6,0.46) [label=above left:$Z$,point];
				\node (c) at (0,0) [label=right :$C$,point];
				
				\path (x) edge (y);
				\path (w) edge (x);
				\path (w) edge (y);
				\path (w) edge (z);
				\path (w) edge (v);
				\path (w) edge (c);
				\path (v) edge (y);
				\path (c) edge (y);
				\path (v) edge (x);
				\path (x) edge (c);
			\end{tikzpicture}
}
\subfloat[]{
	\begin{tikzpicture}
		
		\node (y) at (1.1,-1.22) [label=below right:$Y$,point];
		\node (x) at (-1.1,-1.22) [label=below left:$X$,point];
		\node (w) at (0,1.5) [label=above:$W$,point];
		\node (v) at (1.6,0.46) [label=above right:$V$,point];
		\node (z) at (-1.6,0.46) [label=above left:$Z$,point];
		\node (c) at (0,0) [label=right :$C$,point];
		
		\path (x) edge (y);
		\path (w) edge (x);
		\path (w) edge (y);
		\path (w) edge (z);
		\path (v) edge (w);
		\path (w) edge (c);
		\path (v) edge (y);
		\path (c) edge (y);
		\path (v) edge (x);
		\path (x) edge (c);
	\end{tikzpicture}
}		
\subfloat[]{
	\begin{tikzpicture}
		
		\node (y) at (1.1,-1.22) [label=below right:$Y$,point];
		\node (x) at (-1.1,-1.22) [label=below left:$X$,point];
		\node (w) at (0,1.5) [label=above:$W$,point];
		\node (v) at (1.6,0.46) [label=above right:$V$,point];
		\node (z) at (-1.6,0.46) [label=above left:$Z$,point];
		\node (c) at (0,0) [label=right :$C$,point];
		
		\path (x) edge (y);
		\path (w) edge (x);
		\path (w) edge (y);
		\path (z) edge (w);
		\path (w) edge (v);
		\path (w) edge (c);
		\path (v) edge (y);
		\path (c) edge (y);
		\path (v) edge (x);
		\path (x) edge (c);
	\end{tikzpicture}
}		
}
\end{figure}
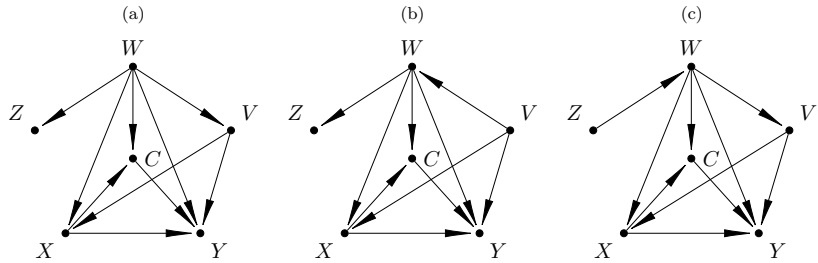

The causal graphs in Figure 5 provide the key exclusion restrictions in this setting.  The strongest restrictions in Figure 5 are on $Z$. $Z$ cannot directly cause or be caused by treatment or outcome.

Suppose $C$ is a post-treatment proxy and $V$ is a pre-treatment proxy, and both are determined prior to the outcome $Y$. In addition, let us assume that $W$ is determined prior to all the observables, with the possible exception of $V$. Then $V$ and $C$ can be caused by all variables determined prior to them other than $Z$ and can determine all variables determined after them other than $Z$.

The exclusion restrictions in Figure 5 are not sufficient for the independence restrictions of the double proxy approach. In the double proxy approach each of the two proxies must be independent of either the treatment or outcome. This effectively rules out any proxies being directly causally related to both $X$ and $Y$. Figure 5 allows $C$ and $V$ to be causally related to both $X$ and $Y$ and so neither can act as proxy in the double proxy case.

\newtheorem*{P4}{Proposition 4}
\begin{P4}
	The NPSEMs associated with the causal graphs in Figure 5 imply the following conditional independence restrictions:
	
	i. $C\indep(V,Z)|(W,X)$, ii. $V\indep Z|(W,X)$, iii. $X\indep Z|W$, iv. $Y\indep Z|(W,V,X)$, and v. $Y(x)\indep X|(W,V)$.
\end{P4}

In this setting Assumption 8 replaces HS Assumption 4.

\theoremstyle{definition}
\newtheorem*{A8}{Assumption 8}
\begin{A8}
	For all $x\in\mathcal{X}$ and any $w_1,w_2\in\mathcal{W}$, if $w_1\neq w_2$ then:
	 \[P\big(f_{C|WX}(C|w_1,x)\neq f_{C|WX}(C|w_2,x)\big)>0\]
\end{A8}

\theoremstyle{plain}
\newtheorem*{T4}{Theorem 3.4 (Auxiliary Proxies)}
\begin{T4}
	Suppose conclusions i., ii., and iii. of Proposition 4 and Assumptions 2, 7, and 8 hold. Then $f_{Z|W}$ is identified up to a reordering of $W$ from the equation below:
	\[
	f_{CZ|VX}(c,z|v,x)=\int_\mathcal{W} f_{C|WX}(c|w,x)f_{W|VX}(w|v,x)f_{Z|W}(z|w)dw
	\]
	In addition, if conclusion iv. of Proposition 4 holds, $f_{YWVX}$ is identified (up to a reordering of $W$) from the linear integral equation below:
	\[
	f_{YZVX}(y,z,v,x)=\int_{\mathcal{W}}f_{YWVX}(y,w,v,x)f_{Z|W}(z|w)dw
	\]
	 Finally, if conclusion v. of Proposition 4 holds then for $\mu_X$-almost all $x_1$:
	\[
		f_{Y(x_1)WX}(y,w,x_2)
		=\int_{\mathcal{V}} f_{Y|WVX}(y|w,v,x_1)f_{VWX}(v,w,x_2)dv
	\]
	 Thus $f_{Y(x)WX}$ is identified for $\mu_X$-almost all $x$, up to a reordering of $W$. Causal objects can then be point identified by say, (\ref{yx1}) or (\ref{yx2}).
\end{T4}

%% file: normalize2.tex
\section{Confounder Effects}

The results in previous section identify causal effects of the treatment $X$, which are invariant to reordering of $W$. This allows us to avoid HS Assumption 5. However, if we do impose this condition then we can point identify a richer set of objects including causal effects of the latent confounders.

Inspired by a result in \citet{Freyberger2021}, we provide a weaker condition than HS Assumption 5 that is sufficient to point identify objects involving the quantile rank of each coordinate of $W$. For example, suppose $W$ is a scalar that represents a student's skill at mathematics. Under a weaker condition than HS Assumption 5 we can identify the causal effect of increasing math skill from the 25-th to the 50-th percentile.

Assumption 9 below is weaker than HS Assumption 5. Rather than assume $M[f_{Z|W}(\cdot|w)]$ is equal to $w$, we require only that it is an unknown  increasing function of $w$. The condition is similar to, but distinct from, the monotonicity assumption in \citet{Freyberger2021} which instead requires that $Z$ can be written as a strictly monotone function of $W$ and some independent noise.

\theoremstyle{definition}
\newtheorem*{AN2}{Assumption 9}
\begin{AN2}
	There is a known functional $M$ so that for some (unknown) function $\phi$, $M[f_{Z|W}(\cdot|w)]=\phi(w),\,\forall w\in\mathcal{W}$.
	$\phi(w)$ has the same length as $W$, each coordinate of $\phi(w)$ depends only on the corresponding coordinate of $w$, and each coordinate is strictly increasing in the corresponding coordinate of $w$.
\end{AN2}

The assumptions above allows us to point identify objects that involve the quantile rank of each coordinate of $W$. Let $Q_W$ be the coordinate-wise quantile function of $W$. That is, for a vector $\tau\in[0,1]^{d_W}$, $Q_W(\tau)$ is the vector whose $k$-th coordinate is the $\tau_k$ quantile of the $k$-th coordinate of $W$, where $d_W$ is the dimension of $W$ and $\tau_k$ is the $k$-th coordinate of $\tau$. Assumption 9 allows us to recover say, the density of $Y(x)$ conditional on $W=Q_W(\tau)$ for a known $\tau$.


\theoremstyle{plain}
\newtheorem*{T41}{Theorem 4.1}
\begin{T41}
Suppose the conditions of any of Theorems 3.1-3.4 hold so that we identify densities $\tilde{f}_{Z|W}$, $\tilde{f}_{Y(x)X|W}$, and $\tilde{f}_{W}$, that are equal to  $f_{Z|W}$, $f_{Y(x)X|W}$, and $\tilde{f}_{W}$ up to a reordering of $W$. Let $\tilde{W}$ be a random variable with density $\tilde{f}_{W}$ and let $\alpha(w)=M[\tilde{f}_{Z|W}(\cdot|w)]$. 

i. Under HS Assumption 5 we have $f_W(w)=f_{\alpha(\tilde{W})}(w)$, and:
\[
f_{Y(x_1)X|W}(y,x_2|w)=\tilde{f}_{Y(x_1)X|W}(y,x_2|\alpha^{-1}(w))
\]

ii. Under Assumption 9:
\[
f_{Y(x_1)X|W}(y,x_2|Q_{W}(\tau))=\tilde{f}_{Y(x_1)X|W}(y,x_2|q(\tau))
\]
Where $q(\tau)=\alpha^{-1}({Q}_{\alpha(\tilde{W})}(\tau))$ for ${Q}_{\alpha(\tilde{W})}$ the coordinate-wise quantile function of $\alpha(\tilde{W})$.
\end{T41}

Suppose treatment is binary and potential outcomes have finite absolute first moments. Under the conditions of Theorem 4.1 and HS Assumption 5 we can identify say, the average treatment effect within a given stratum $w$ of $W$:\footnote{Strictly speaking we identify this object for almost all $w$.}
\[
E[Y(1)-Y(0)|W=w]
\]
While Assumption 9 is insufficient to identify the object above, Theorem 4.1.ii shows this condition it does allow us to identify say, the average treatment effect within the stratum of $W$ with quantile rank $\tau$:
\[
E[Y(1)-Y(0)|W=Q_W(\tau)]
\]
For example, we can identify the average treatment effect for individuals with ability in the $25$th percentile.

We can use similar arguments to identify causal effects of the confounders $W$ themselves. Let $Y(x,w)$ denote the potential outcome under a counterfactual in which $X$ and $W$ are respectively set to values $x$ and $w$. Proposition 5 provides conditional independence restrictions involving $Y(x,w)$ which follow from the causal graphs in the previous section. These independence conditions enable us to identify causal effects of the latent factors themselves.

\newtheorem*{P5}{Proposition 5}
\begin{P5}
	The NPSEMs associated with all of the causal graphs in Figures 2, 3.a, 3.b, and 4 imply i. $Y(x,w)\indep (X,W)$. The NPSEMs associated with the graphs in Figure 3.c and Figure 6 imply ii. $Y(x,w)\indep (X,W)|V$.
\end{P5}

Under the conditions of any of the theorems in Section 3, and one of the conclusions of Proposition 5, we can identify the joint distribution of $Y(x,w)$ and $X$ up to a reordering of $W$.

\newtheorem*{L1}{Lemma 4.1}
\begin{L1}
	Suppose the conditions of any of Theorems 1, 2, 3 and conclusion i. of Proposition 5 holds, or the conditions of Theorem 4 and conclusion ii. of Proposition 5 hold. Then for $\mu_X$-almost all $x$ and almost all $w$,  $f_{Y(x,w)X}$, $f_{Z|W}$, and $f_W$ are identified up to a reordering of $W$.
\end{L1}

\theoremstyle{plain}
\newtheorem*{T42}{Theorem 4.2}
\begin{T42}
	Suppose the conditions of any of Lemma 4.1 hold so that we identify densities $\tilde{f}_{Z|W}$, $\tilde{f}_{Y(x,w)X}$, and $\tilde{f}_W$ that are equal to $f_{Z|W}$, $f_{Y(x,w)X}$, and ${f}_W$ up to a reordering of $W$. Define $\alpha$ and $q$ as in Theorem 4.1. 
	
	i. Under HS Assumption 5:
	\[
	f_{Y(x_1,w)X}(y,x_2)=\tilde{f}_{Y(x_1,\alpha^{-1}(w))X}(y,x_2)
	\]
	
	ii. Under Assumption 9:
	\[
	f_{Y(x_1,Q_{W}(\tau))X}(y,x_1)=\tilde{f}_{Y(x,q(\tau))X}(y,x_2)
	\]
\end{T42}

If $Y(x,w)$ has bounded derivatives then Theorem 4.2.i allows us to identify say, the average partial effect of an increase in $W$:
\[E[\frac{\partial}{\partial w}Y(x,w)]\]

In the context of the educational example, the above isolates the contribution of academic aptitude to a student's GPA. Theorem 4.2.ii identifies the average partial effect of an increase in the quantile rank of $W$:
\[E\big[\frac{\partial}{\partial \tau}Y\big(x,Q_W(\tau)\big)\big]\]

The object above is likely to be of only limited use when designing an optimal intervention on $W$. This is because it is uninformative about the size of a change in $W$ required to achieve a given causal impact. However, if $W$ measures ability then we can never hope to design such an intervention and the quantile rank of $W$ may be more readily interpretable than the level of $W$ (see \cite{Freyberger2021} for discussion).

	

%% file: conclusion.tex
\section{Conclusion and Further Comparison with Double Proxies}

In this work we establish identification of causal objects using the `triple proxy' approach. We show that there are sets of exclusion restrictions under which we can establish identification (or partial identification) using the triple proxy approach, but not using double proxies. Conversely, there are exclusion restrictions that support the double proxy but not the triple proxy approach. In some settings the exclusion restrictions may allow for both approaches, in this case a  comparison of the merits of the two strategies is more nuanced.

One advantage of the triple proxy approach is that it identifies some objects  not identified using the double proxy approach. In particular, objects that measure the degree of  heterogeneity in treatment effects between strata of the latent variables. Moreover, under some additional conditions, we can adapt our approach to identify  causal effects of the latent variables themselves.

An important distinction is that the double proxy approach identifies causal effects from equations involving only observables. This avoids the need to directly specify any modeling assumptions on distributions of the latent factors and simplifies estimation. However, we may wish to impose a priori constraints on the densities of latent factors either to improve precision, or to test these conditions. The double proxy approach precludes this. By contrast, the triple proxy approach is built on equations involving densities of the latent confounders and so we could impose such constraints in estimation.

Simple non-parametric estimators of causal effects are available for the double proxy approach. For example, \citet{Deaner2021} suggests an estimator that is similar to sieve two-stage least squares. We leave nonparametric estimation using the triple proxy approach as an open problem. \citet{Hu2008} suggest a sieve maximum likelihood method which one could apply to the HS step in our identification results. However, for some choices of the third proxy and conditioning variables, our identification strategy involves a second integral equation. One may be able to apply NPIV methods to estimate a solution to this equation. In the meantime the nonparametric identification results presented here may act as motivation for a parametric estimation strategy or estimation of a discretized version of the model.

%% file: partial.tex
\appendix

\section{Weakening the Exclusion Restrictions Using Rank Invariance}

In this section we consider a particular rank invariance condition involving the conditional (on $W$) average treatment effect (CATE). More precisely, we assume that individuals in strata of $W$ with higher untreated average potential outcomes have larger average treatment effects. We show that under this condition, we can partially identify conditional and unconditional average treatment effects under weaker exclusion restrictions than those in Section 3.

More precisely, we are able to weaken the exclusion restrictions in the outcome proxy and auxiliary proxy cases explored in Sections 3.1 and 3.4. When the CATE is constant we achieve point identification.

We assume throughout that treatment is binary with $X=1$ indicating treatment and $X=0$ no treatment. However, the results extend straight-forwardly to more general discrete treatments.

For some intuition, recall that the identification results in Sections 3.1 and 3.4 require that $Z$ not cause or be caused by $X$. The need for this exclusion restriction arises because we apply \citet{Hu2008} within each stratum of $X$ and identify objects up to reorderings of $W$. The exclusion restriction helps to ensure that the reorderings do not differ between strata of the treatment. Rank invariance allows us to make some limited comparison of the conditional average potential outcomes between treated and untreated individuals even when the ordering of $W$ varies with treatment status.

\subsection{Outcome Proxies with Rank Invariance}

We apply \citet{Hu2008} with $C=Y$ as in Section 3.1. We weaken the exclusion restrictions in Figure 2 to those in Figure 6.

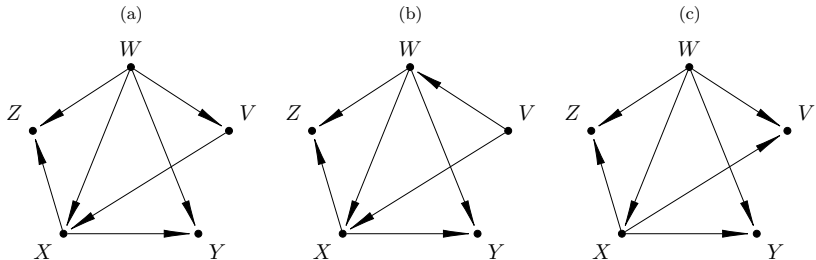
\begin{figure}[h]
	\centering
	\caption{Outcome Proxy Graph}
	
	\centering
	\resizebox{!}{11pt}{%
		\subfloat[]{
			
			\begin{tikzpicture}
				
				\node (y) at (1.1,-1.22) [label=below right:$Y$,point];
				\node (x) at (-1.1,-1.22) [label=below left:$X$,point];
				\node (w) at (0,1.5) [label=above:$W$,point];
				\node (v) at (1.6,0.46) [label=above right:$V$,point];
				\node (z) at (-1.6,0.46) [label=above left:$Z$,point];
				
				\path (x) edge (y);
				\path (w) edge (x);
				\path (w) edge (y);
				\path (w) edge (z);
				\path (w) edge (v);
				\path (v) edge (x);
				\path (x) edge (z);
				
			\end{tikzpicture}
		}
		\subfloat[]{
			\begin{tikzpicture}
				
				\node (y) at (1.1,-1.22) [label=below right:$Y$,point];
				\node (x) at (-1.1,-1.22) [label=below left:$X$,point];
				\node (w) at (0,1.5) [label=above:$W$,point];
				\node (v) at (1.6,0.46) [label=above right:$V$,point];
				\node (z) at (-1.6,0.46) [label=above left:$Z$,point];
				
				\path (x) edge (y);
				\path (w) edge (x);
				\path (w) edge (y);
				\path (w) edge (z);
				\path (v) edge (w);
				\path (v) edge (x);
				\path (x) edge (z);
				
			\end{tikzpicture}
		}
		\subfloat[]{
			
			\begin{tikzpicture}
				
				\node (y) at (1.1,-1.22) [label=below right:$Y$,point];
				\node (x) at (-1.1,-1.22) [label=below left:$X$,point];
				\node (w) at (0,1.5) [label=above:$W$,point];
				\node (v) at (1.6,0.46) [label=above right:$V$,point];
				\node (z) at (-1.6,0.46) [label=above left:$Z$,point];
				
				\path (x) edge (y);
				\path (w) edge (x);
				\path (w) edge (y);
				\path (w) edge (z);
				\path (w) edge (v);
				\path (x) edge (v);
				\path (x) edge (z);
				
			\end{tikzpicture}
		}
	}	
\end{figure}

The graphs in Figure 6 weaken those in Figure 2 by allowing treatment $X$ to directly impact $Z$ which is a vector of post-treatment proxies. Consider Figure 6.a, in the test score case, $V$ is a vector of pre-treatment test scores that can directly determine treatment and $Z$ is a vector of post-treatment scores which can be directly affected by treatment.

The restrictions in Figure 6 are not sufficiently strong for the double proxy approach. Figure 6 allows treatment to be directly causally related to both the proxies $Z$ and $V$, which is incompatible with the double proxy approach. 

\theoremstyle{plain}
\newtheorem*{P6}{Proposition 6}
\begin{P6}
	The NPSEMs associated with the causal graphs in Figure 6 all imply the following conditional independence restrictions:
	
	i. $Y\indep(V,Z)|(W,X)$, ii. $V\indep Z|(W,X)$, and iii. $Y(x)\indep (X,V)|W$
\end{P6}

The conclusions of Proposition 6 are weaker than those of Proposition 1. In particular, we drop conclusion iii. of Proposition 1 (independence of $X$ and $Z$ conditional on $W$). In contrast to the double proxy approach, $V$ is not required to be independent of $X$ conditional on any of the other variables.

In order to partially identify conditional average treatment effects we require a rank invariance assumption given below.

\theoremstyle{definition}
\newtheorem*{A10}{Assumption 10 (Rank invariance)}
\begin{A10}
	There is a constant $c<\infty$ so that $|E[Y(0)|W]|$ is almost surely bounded by $c$. Moreover, for any $w_1,w_2\in\mathcal{W}$, if $E[Y(0)|W=w_2]\geq E[Y(0)|W=w_1]$ then:
	\[
	E[Y(1)-Y(0)|W=w_2]\geq E[Y(1)-Y(0)|W=w_1]
	\]
\end{A10}

Assumption 10 states that if the average untreated outcome is larger in one stratum of $W$ than another, then the average treatment effect in that stratum is also larger. If a large value of $Y$ indicates a more favorable outcome, then loosely speaking, those who do better without treatment tend to benefit more from treatment.

Note that the inequalities need not be strict. Assumption 10 allows for the possibility that  $E[Y(1)-Y(0)|W=w]$ is constant for all $w$.

\theoremstyle{plain}
\newtheorem*{T7}{Theorem A.1 (monotone CATE with outcome proxies)}
\begin{T7}
Suppose the conclusions of Proposition 6 holds and Assumption 1, 2, and 3, hold. Then for $x=1,2$, ${f}_{Y(x)|W}$ is identified up to reorderings of $W$ which may depend on $x$.

Thus we identify functions $\tilde{f}_{Y(1)|W}$ and $\tilde{f}_{Y(0)|W}$ which differ from ${f}_{Y(1)|W}$ and ${f}_{Y(0)|W}$ in the orderings of $W$. Define  $\bar{s}$ and $\underline{s}$ as follows:
\begin{align*}
\bar{s}&=\underset{w\in\mathcal{W}}{\text{ess sup}}\int_{\mathcal{Y}}y\tilde{f}_{Y(1)|W}(y|w)dy-\underset{w\in\mathcal{W}}{\text{ess sup}}\int_{\mathcal{Y}}y\tilde{f}_{Y(0)|W}(y|w)dy\\
\underline{s}&	=\underset{w\in\mathcal{W}}{\text{ess inf}}\int_{\mathcal{Y}}y\tilde{f}_{Y(1)|W}(y|w)dy-\underset{w\in\mathcal{W}}{\text{ess inf}}\int_{\mathcal{Y}}y\tilde{f}_{Y(0)|W}(y|w)dy
\end{align*}
Under Assumption 10 $\bar{s}$ and $\underline{s}$ are respectively the essential supremum and infimum of $E[Y(1)-Y(0)|W=w]$ and for $x=0,1$:
\[
E[Y(1)-Y(0)|X=x]\in[\underline{s},\bar{s}]
\]
\end{T7}

Theorem 6 partially identifies the CATE, the average effect of treatment on the treated, and the average effect of treatment on the untreated. If the CATE is constant then $\underline{s}=\bar{s}$ and so the identified set is a singleton and the effects are point identified.

\subsection{Auxiliary Proxies with Rank Invariance}

Finally we revisit the case examined in Section 3.4 in which $C$ is a vector of additional variables. We weaken the exclusion restrictions in Figure 5 to those in Figure 7.

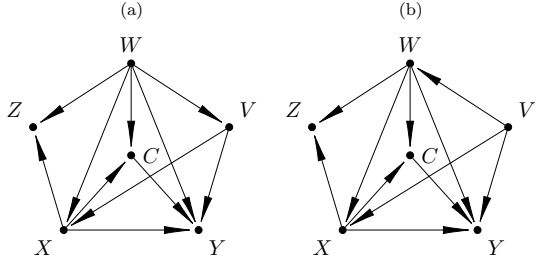
\begin{figure}[h]
	\centering
	\caption{Auxiliary Proxy Graphs}
	
	\centering	
	\resizebox{!}{11pt}{	
		\subfloat[]{
			\begin{tikzpicture}
				
				\node (y) at (1.1,-1.22) [label=below right:$Y$,point];
				\node (x) at (-1.1,-1.22) [label=below left:$X$,point];
				\node (w) at (0,1.5) [label=above:$W$,point];
				\node (v) at (1.6,0.46) [label=above right:$V$,point];
				\node (z) at (-1.6,0.46) [label=above left:$Z$,point];
				\node (c) at (0,0) [label=right :$C$,point];
				
				\path (x) edge (y);
				\path (w) edge (x);
				\path (w) edge (y);
				\path (w) edge (z);
				\path (w) edge (v);
				\path (w) edge (c);
				\path (v) edge (y);
				\path (c) edge (y);
				\path (x) edge (z);
				\path (v) edge (x);
				\path (x) edge (c);
			\end{tikzpicture}
		}
		\subfloat[]{
			\begin{tikzpicture}
				
				\node (y) at (1.1,-1.22) [label=below right:$Y$,point];
				\node (x) at (-1.1,-1.22) [label=below left:$X$,point];
				\node (w) at (0,1.5) [label=above:$W$,point];
				\node (v) at (1.6,0.46) [label=above right:$V$,point];
				\node (z) at (-1.6,0.46) [label=above left:$Z$,point];
				\node (c) at (0,0) [label=right :$C$,point];
				
				\path (x) edge (y);
				\path (w) edge (x);
				\path (w) edge (y);
				\path (w) edge (z);
				\path (v) edge (w);
				\path (w) edge (c);
				\path (v) edge (y);
				\path (c) edge (y);
				\path (v) edge (x);
					\path (x) edge (z);
				\path (x) edge (c);
			\end{tikzpicture}
		}			
	}
\end{figure}

The causal graphs in Figure 7 differ from those in Figures 6.a and 6.b in that they allow $Z$ to be a post-treatment variable that may be directly affected treatment. Note that $Z$ still cannot directly affect the outcome.

Note that we do not include a relaxed version on Figure 6.c. If we relaxed 7.c to allow $X$ to affect $Z$ the resulting graph would be cyclic.

The graphs in Figure 7 imply conditional independence restrictions given in Proposition 7. These conditions weaken those in Proposition 4. In particular we drop condition iv. of  the proposition and leave the remaining conditions unchanged.

\newtheorem*{P7}{Proposition 7}
\begin{P7}
	The NPSEMs associated with the causal graphs in Figure 7 imply the following conditional independence restrictions:
	
	i. $C\indep(V,Z)|(W,X)$, ii. $V\indep Z|(W,X)$, iii. $Y\indep Z|(W,V,X)$, and iv. $Y(x)\indep X|(W,V)$.
\end{P7}

In this setting we adapt Assumption 10 to apply within each stratum of $V$. We state this condition as Assumption 11 below.

\theoremstyle{definition}
\newtheorem*{A11}{Assumption 11 ($V$-conditional rank invariance)}
\begin{A11}
	For almost all $v\in\mathcal{V}$ there is a constant $c<\infty$ so that $|E[Y(0)|W,V=v]|<c$ with probability $1$. Moreover, for any $w_1,w_2\in\mathcal{W}$, if $E[Y(0)|W=w_2,V=v]\geq E[Y(0)|W=w_1,V=v]$ then:
	\[
	E[Y(1)-Y(0)|W=w_2,V=v]\geq E[Y(1)-Y(0)|W=w_1,V=v]
	\]
\end{A11}

As in the outcome proxy case, rank invariance allows us to partially identify causal effects. the identified set reduces to a point when for almost all $v\in\mathcal{V}$, $E[Y(1)-Y(0)|W=w,V=v]$ does not depend on $w$.

\theoremstyle{plain}
\newtheorem*{T8}{Theorem A.2 (Rank invariance with auxiliary proxies)}
\begin{T8}
	Suppose conclusions of Proposition 7 and Assumptions 2, 7, and 8 hold.  Then for $x=1,2$, ${f}_{Y(x)|WV}$ is identified up to reorderings of $W$ which may depend on $x$.

Thus we identify functions $\tilde{f}_{Y(1)|WV}$ and $\tilde{f}_{Y(0)|WV}$ which differ from ${f}_{Y(1)|WV}$ and ${f}_{Y(0)|WV}$ in the orderings of $W$. Define  $\bar{s}(v)$ and $\underline{s}(v)$ as follows:
\begin{align*}
	\bar{s}(v)&=\underset{w\in\mathcal{W}}{\text{ess sup}}\int_{\mathcal{Y}}y\tilde{f}_{Y(1)|WV}(y|w,v)dy-\underset{w\in\mathcal{W}}{\text{ess sup}}\int_{\mathcal{Y}}y\tilde{f}_{Y(0)|WV}(y|w,v)dy\\
	\underline{s}(v)&	=\underset{w\in\mathcal{W}}{\text{ess inf}}\int_{\mathcal{Y}}y\tilde{f}_{Y(1)|WV}(y|w,v)dy-\underset{w\in\mathcal{W}}{\text{ess inf}}\int_{\mathcal{Y}}y\tilde{f}_{Y(0)|WV}(y|w,v)dy
\end{align*}
Under Assumption 11 for almost all $v\in\mathcal{V}$, $\bar{s}(v)$ and $\underline{s}(v)$ are respectively the essential supremum and infimum over $w$ of $E[Y(1)-Y(0)|W=w,V=v]$ and for $x=0,1$:
\[E[Y(1)-Y(0)|X=x]\in\big[E[\underline{s}(V)|X=x],E[\bar{s}(V)|X=x]\big]
\]
\end{T8}

%% file: proofs.tex
\appendix

\section{Proofs}

\begin{proof}[Proof of Theorem 3.1] Under conditions i. and ii.
	of Proposition 1 and Assumptions 1, 2, and 3, we can apply HS Theorem
	1 conditional on $X=x_{1}$ with $C=Y$. This yields identification
	of $f_{Y|WX}(\cdot|\cdot,x_{1})$, $f_{Z|WX}(\cdot|\cdot,x_{1})$,
	and $f_{W|VX}(\cdot|\cdot,x_{1})$ up to a reordering of $W$ from
	the equation below: 
	\begin{align}
		f_{YZ|VX}(y,z|v,x_{1}) & =\int_{\mathcal{W}}f_{Y|WX}(y|w,x_{1})f_{Z|W}(z|w)f_{W|VX}(w|v,x_{1})dw\label{eq:id11}
	\end{align}
	In the above we have imposed that $f_{Z|W}(z|w)=f_{Z|WX}(z|w,x_{1})$
	which follows from conclusion iii. of Proposition 1. Now, we could
	apply the same reasoning for some other value $x_{2}$ of $X$, however
	we need to ensure we get the same ordering as with $x_{1}$. Now,
	taking (\ref{eq:id11}) with $x_{2}$ in place of $x_{1}$, and integrating
	over $y\in\mathcal{\mathcal{Y}}$ we get:
	
	\begin{align*}
		f_{Z|VX}(z|v,x_{2}) & =\int_{\mathcal{W}}f_{Z|W}(z|w)f_{W|VX}(w|v,x_{2})dw
	\end{align*}
	
	We have already identified $f_{Z|W}$, and $f_{Z|VX}$ only involves
	observables, so the only unknown in the above is $f_{W|VX}(\cdot|\cdot,x_{2})$.
	By Assumption 2, for each $v$ the above admits a unique solution
	$f_{W|VX}(\cdot|\cdot,x_{2})$. To see this, suppose that for a given
	$v$ there are two solutions $h_{1}$ and $h_{2}$, both of which
	are bounded and integrable. Then we must have:
	
	\[
	\int_{\mathcal{W}}f_{Z|W}(z|w)\big(h_{1}(w)-h_{2}(w)\big)dw=0
	\]
	But then we apply Assumption 2 with $\delta(w)=h_{1}(w)-h_{2}(w)$
	and we see that $h_{1}(w)=h_{2}(w)$. Thus (\ref{eq:id11}) identifies
	$f_{W|VX}(\cdot|\cdot,x_{2})$ up to reorderings of $W$. By similar
	reasoning we get that $f_{Y|WX}(\cdot|\cdot,x_{2})$ is then identified
	from \ref{eq:id11} with $x_{1}$ replaced by $x_{2}$ up to reorderings
	of $W$. Since $x_{2}$ was chosen arbitrarily, we get that $f_{W|VX}$
	and $f_{Y|WX}$ are identified up to a reordering of $W$.
	
	Now, by elementary properties of probability densities we have:
	
	\[
	f_{WX}(w,x)=\int_{\mathcal{V}}f_{W|VX}(w|v,x)f_{VX}(v,x)dv
	\]
	The objects on the RHS above are all known (up to a reordering of
	$W$), and so $f_{WX}$ is identified. Now note that by Assumption
	1, conclusion iv. of Proposition 1, and consistency $Y(X)=Y$, for
	$\mu_{X}$-almost all $X$: 
	\begin{align*}
		f_{Y(x_{1})|WX}(y|w,x_{2})= & f_{Y(x_{1})|WX}(y|w,x_{1})\\
		= & f_{Y|WX}(y|w,x_{1})
	\end{align*}
	
	In all, $f_{Y(x)|WX}$ is identified, for $\mu_{X}$-almost all $x$,
	up to a reordering of $W$. Multiplying by $f_{WX}$ gives the result.
\end{proof}

\begin{proof}[Proof of Theorem 3.2]
	
	Under conditions i. and ii. of Proposition 2, HS Assumption 3, Assumption
	1, and Assumption 4, we can apply HS Theorem 1 with $C=X$. This yields
	identification of $f_{X|W}$, $f_{W|V}$, and $f_{Z|W}$ up to a reordering
	of $W$ from the equation in the theorem. Now note that by elementary
	properties of probability densities: 
	\begin{align*}
		f_{YZX}(y,z,x) & =\int_{\mathcal{W}}f_{YWZX}(y,w,z,x)dw\\
		& =\int_{\mathcal{W}}f_{Y|XWZ}(y|x,w,z)f_{WZX}(w,z,x)dw\\
		& =\int_{\mathcal{W}}f_{Y|XWZ}(y|x,w,z)f_{Z|WX}(z|w,x)f_{XW}(x,w)dw
	\end{align*}
	
	By conclusion ii of Proposition 2, $f_{Z|XW}(z|x,w)=f_{Z|W}(z|w)$.
	Condition iii. of Proposition 2 implies $f_{Y|XWZ}(y|x,w,z)=f_{Y|XW}(y|x,w)$.
	Substituting into the equation above and simplifying we get: 
	\begin{align*}
		f_{YZX}(y,z,x) & =\int_{\mathcal{W}}f_{Y|XW}(y|x,w)f_{Z|W}(z|w)f_{XW}(x,w)dw\\
		& =\int_{\mathcal{W}}f_{YXW}(y,x,w)f_{Z|W}(z|w)dw
	\end{align*}
	Other than $f_{YWX}$, all the objects in the above are already identified
	up to reorderings of $W$. To show $f_{YWX}$ is the unique solution
	to the equation, note that by HS Assumption 3 there can be only one
	bounded integrable function $h$ so that for all $z\in\mathcal{Z}$:
	\[
	f_{YZX}(y,z,x)=\int_{\mathcal{W}}h(w)f_{Z|W}(z|w)dw
	\]
	If there were two such functions, $h_{1}$ and $h_{2}$ then we would
	have: 
	\[
	\int_{\mathcal{W}}\big(h_{1}(w)-h_{2}(w)\big)f_{Z|W}(z|w)dw=0,\,\forall z\in\mathcal{Z}
	\]
	and so by HS Assumption 3, $h_{1}(W)-h_{2}(W)=0$ almost surely. Therefore,
	the only solution is: 
	\[
	h(w)=f_{YWX}(y,w,x)
	\]
	Thus $f_{YWX}$ is identified up to a reordering of $W$. Under conclusion
	iv. of Proposition 2 we can now apply the final steps in the proof
	of Theorem 1 to identify $f_{Y(x)WX}$ up to a reordering of $W$.
\end{proof}

\begin{proof}[Proof of Theorem 3.3] Under conditions i. and ii.
	of Proposition 3 and Assumption 1, 5, and 6, we can apply HS Theorem
	1 conditional on $Y=y$ with $C=X$. This yields identification of
	$f_{X|WY}(\cdot|\cdot,y)$, $f_{Z|WY}(\cdot|\cdot,y)$, and $f_{W|VY}(\cdot|\cdot,y)$
	(up to a reordering of $W$) from the equation below: 
	\begin{align}
		f_{XZ|VY}(x,z|v,y) & =\int_{\mathcal{W}}f_{X|WY}(x|w,y)f_{W|VY}(w|v,y)f_{Z|WY}(z|w,y)dw\label{eq:id11-1}
	\end{align}
	
	In the above we have imposed condition iii. of Proposition 3 which
	implies that $f_{Z|WY}(z|w,y)=f_{Z|W}(z|w)$. Taking (\ref{eq:id11-1})
	with $y'$ in place of $y$ and integrating over $x\in\mathcal{\mathcal{X}}$
	we get:
	
	\begin{align*}
		f_{Z|VY}(z|v,y') & =\int_{\mathcal{W}}f_{Z|W}(z|w)f_{W|VY}(w|v,y')dw
	\end{align*}
	Other than $f_{W|VY}(\cdot|\cdot,y')$, all the objects in the equation
	above are identified (at least up to a reordering of $W$). By Assumption
	5, for each $z$ the equation above has a unique solution $f_{W|VY}(\cdot|\cdot,y')$
	(see the reasoning in the proof of Theorem 1). Thus we have identified
	$f_{W|VY}$ up to a reordering of $W$ from (\ref{eq:id11-1}). By
	similar reasoning $f_{X|WY}(x|w,y')$ is identified from (\ref{eq:id11-1})
	up to a reordering of $W$. Now, by elementary properties of densities:
	\[
	f_{YXW}(y,x,w)=f_{X|WY}(x|w,y)\int f_{W|VY}(w|v,y)f_{VY}(v,y)dv
	\]
	Since the densities on the RHS are all identified up to a reordering
	of $W$, $f_{YXW}$ is identified up to a reordering of $W$ and thus
	so are $f_{Y|XW}$ and $f_{XW}$. Applying conclusion iv. of Proposition
	3 and following the same steps as in the proof of Theorem 1, we identify
	$f_{Y(x)WX}$ up to a reordering of $W$. \end{proof}

\begin{proof}[Proof of Theorem 3.4]
	Under conclusions i. and ii.
	of Proposition 4 and Assumptions 2, 7, and 8 we can apply HS Theorem
	1 within the stratum $x$ of $X$. This yields identification of $f_{C|WX}(\cdot|\cdot,x)$,
	$f_{W|VX}(\cdot|\cdot,x)$, and $f_{Z|WX}(\cdot|\cdot,x)$ up to a
	reordering of $W$ from the equation below: 
	\[
	f_{CZ|VX}(c,z|v,x_{1})=\int_{\mathcal{W}}f_{C|WX}(c|w,x_{1})f_{W|VX}(w|v,x_{1})f_{Z|W}(z|w)dw
	\]
	Note we have imposed $f_{Z|WX}(z|w,x)=f_{Z|W}(z|w)$ which follows
	from conclusion iii. of Proposition 4.
	
	Now, by elementary properties of probability densities we have: 
	\begin{align*}
		&f_{YZVX}(y,z,v,x) \\ =&\int_{\mathcal{W}}f_{Y|WVZX}(y|w,v,z,x)f_{Z|WVX}(z|w,v,x)f_{WVX}(w,v,x)dw
	\end{align*}
	Proposition 4.iv implies $f_{Y|WVZX}(y|w,v,z,x)=f_{Y|WVX}(y|w,v,x)$.
	In addition, conclusions ii. and iii. of Proposition 4 imply $(V,X)\indep Z|W$
	and so $f_{Z|WVX}(z|w,v,x)=f_{Z|W}(z|w)$. Substituting and again
	applying elementary properties of probability densities we get: 
	\begin{align*}
		f_{YZVX}(y,z,v,x) & =\int_{\mathcal{W}}f_{Y|WVX}(y|w,v,x)f_{Z|W}(z|w)f_{WVX}(w,v,x)dw\\
		& =\int_{\mathcal{W}}f_{YWVX}(y,w,v,x)f_{Z|W}(z|w)dw
	\end{align*}
	
	All objects in the equation above are identified up to a reordering
	of $W$, other than $f_{YWVX}$. By the same reasoning as in the proof
	of Theorem 1, Assumption 2 implies there is only one bounded integrable
	function $h$ so that for all $z\in\mathcal{Z}$: 
	\[
	f_{YVZX}(y,v,z,x)=\int_{\mathcal{W}}h(w)f_{Z|W}(z|w)dw
	\]
	The only solution to the above is: 
	\[
	h(w)=f_{YWVX}(y,w,v,x)
	\]
	Thus we have identified $f_{YWVX}$ up to a reordering of $W$. Finally,
	using conclusion v. of Proposition 4 we identify $f_{Y(x)WX}$ for
	$\mu_{X}$-almost all $x$ up to a reordering of $W$: 
	\begin{align*}
		f_{Y(x_{1})WX}(y,w,x_{2}) & =\int_{\mathcal{V}}f_{Y(x_{1})|WVX}(y|w,v,x_{2})f_{VWX}(v,w,x_{2})dv\\
		& =\int_{\mathcal{V}}f_{Y(x_{1})|WVX}(y|w,v,x_{1})f_{VWX}(v,w,x_{2})dv\\
		& =\int_{\mathcal{V}}f_{Y|WVX}(y|w,v,x_{1})f_{VWX}(v,w,x_{2})dv
	\end{align*}
	
	Where the second equality uses conclusion v. of Proposition 4 and
	the third uses consistency ($Y(X)=Y$).
	\end{proof}
	
	\begin{proof}[Proof of Lemma 4.1]
		By consistency, for $\mu_{X}$-almost all $x_{1}$ and almost all
		$w$, $f_{Y|XW}(y|x_{1},w)=f_{Y(x_{1},w)|XW}(y|x_{1},w)$. By conclusion
		i. of Proposition 5, $f_{Y(x_{1},w)|XW}(y|x_{1},w)=f_{Y(x_{1},w)|XW}(y|x_{2},w_{2})$.
		Combining we get: 
		\[
		f_{Y(x_{1},w)X}(y,x_{2})=\int_{\mathcal{W}}f_{Y|XW}(y|x_{1},w)f_{WX}(w_{2},x_{2})dw_{2}
		\]
		In each of Theorems 1-3 we identify $f_{Y|XW}$ and $f_{WX}$ up to
		a reordering of $W$ as intermediate objects and thus $f_{Y(x,w)X}$
		is identified up to reorderings of $W$ under 5.i and the conditions
		of any of these theorems.
		
		Also by consistency, for $\mu_{X}$-almost all $x_{1}$ and almost
		all $w$, $f_{Y|XVW}(y|x_{1},v,w)=f_{Y(x_{1},w)|XVW}(y|x_{1},v,w)$,
		and by conclusion ii. of Proposition 5: \[f_{Y(x_{1},w)|XVW}(y|x_{1},v,w)=f_{Y(x_{1},w)|XVW}(y|x_{2},v,w_{2})\]
		Combining we get: 
		\[
		f_{Y(x_{1},w)X}(y,x_{2})=\int_{\mathcal{V}}\int_{\mathcal{W}}f_{Y|XVW}(y|x_{1},v,w)f_{WVX}(w_{2},v,x_{2})dw_{2}dv
		\]
		In Theorem 4 we identify $f_{Y|XVW}$ and $f_{WVX}$ up to a reordering
		of $W$ as intermediate objects. That is, we identify $f_{Y|XV\varphi(W)}$
		and $f_{\varphi(W)VX}$. Thus under the conditions of this theorem
		and 5.ii we identify $f_{Y(x_{1},w)X}$ up to a reordering of $W$.
	\end{proof}
	
	\begin{proof}[Proof of Theorem 4.1]
		By any of Theorems 1, 2, 3,
		or 4 we achieve identification of $f_{Y(x)X|W}$, $f_{W}$, and $f_{Z|W}$
		up to a reordering of $W$. More precisely, $\tilde{f}_{Y(x)WX}$,
		$\tilde{f}_{Z|W}$, and $\tilde{f}_{W}$ are identified, where $\tilde{f}_{Y(x_{1})X|W}(y,x_{2}|w)=f_{Y(x_{1})X|\varphi(W)}(y,x_{2}|w)$,
		$\tilde{f}_{Z|W}(z|w)=f_{Z|\varphi(W)}(z|w)$, and $\tilde{f}_{W}(w)=f_{\varphi(W)}(w)$
		for an unknown injective function $\varphi$. From $\tilde{f}_{W}(w)=f_{\varphi(W)}(w)$
		we see that $\tilde{W}=\varphi(W)$. Then we have:
		\begin{align*}
			\alpha(w) & =M[\tilde{f}_{Z|W}(\cdot|w)]\\
			& =M[f_{Z|\varphi(W)}(\cdot|w)]\\
			& =M[f_{Z|W}(\cdot|\varphi^{-1}(w))]
		\end{align*}
		
		Under HS Assumption 5, $M[f_{Z|W}(\cdot|w)]=w$, and so $\alpha(w)=\varphi^{-1}(w)$.
		Thus $\alpha(\tilde{W})=W$ and so $f_{W}(w)$ is equal to $f_{\alpha(\tilde{W})}(w)$.
		Moreover, we have:
		\begin{align*}
			f_{Y(x_{1})X|W}(y,x_{2}|w) & =f_{Y(x_{1})X|W}(y,x_{2}|w)\\
			& =f_{Y(x_{1})X|\alpha(\varphi(W))}(y,x_{2}|w)\\
			& =f_{Y(x_{1})X|\varphi(W)}(y,x_{2}|\alpha^{-1}(w))\\
			& =\tilde{f}_{Y(x_{1})X|W}(y,x_{2}|\alpha^{-1}(w))
		\end{align*}
		
		Under Assumption 9, which is weaker than HS Assumption 5, there is
		a component-wise strictly increasing function $\phi$ so that $M[f_{Z|W}(\cdot|w)]=\phi(w)$,
		and so: 
		\[
		\alpha(w)=\phi(\varphi^{-1}(w))
		\]
		
		It follows that $\alpha(\tilde{W})=\phi(W)$. Since each coordinate
		of $\phi(w)$ is a strictly increasing function of the corresponding
		coordinate of $w$, $Q_{\phi(W)}(\tau)=\phi\big(Q_{W}(\tau)\big)$.
		Thus we have $Q_{\alpha(\tilde{W})}(\tau)=\phi\big(Q_{W}(\tau)\big)$
		and so $\phi^{-1}\big(Q_{\alpha(\tilde{W})}(\tau)\big)=Q_{W}(\tau)$.
		Putting this together we get:
		\begin{align*}
			\tilde{f}_{Y(x_{1})X|W}\big(y,x_{2}|\alpha^{-1}(Q_{\alpha(\tilde{W})}(\tau))\big) & =f_{Y(x_{1})X|\varphi(W)}\big(y,x_{2}|\alpha^{-1}(Q_{\alpha(\tilde{W})}(\tau))\big)\\
			& =f_{Y(x_{1})X|\alpha(\varphi(W))}\big(y,x_{2}|Q_{\alpha(\tilde{W})}(\tau)\big)\\
			& =f_{Y(x_{1})X|\phi(W)}\big(y,x_{2}|Q_{\alpha(\tilde{W})}(\tau)\big)\\
			& =f_{Y(x_{1})X|W}\big(y,x_{2}|\phi^{-1}(Q_{\alpha(\tilde{W})}(\tau))\big)\\
			& =f_{Y(x_{1})X|W}(y,x_{2}|Q_{W}(\tau))
		\end{align*}
		As required.
			\end{proof}
	
	\begin{proof}[Proof of Theorem 4.2]
		 From Lemma 1 we achieve identification
		of $f_{Y(x,w)X}$, $f_{W}$, and $f_{Z|W}$ up to a reordering of
		$W$. More precisely, $\tilde{f}_{Y(x,w)X}$, $\tilde{f}_{Z|W}$,
		and $\tilde{f}_{W}$ are identified, where $\tilde{f}_{Y(x_{1},w)X}(y,x_{2})=f_{Y(x_{1},\varphi^{-1}(W))X}(y,x_{2})$,
		$\tilde{f}_{Z|W}(z|w)=f_{Z|\varphi(W)}(z|w)$, and $\tilde{f}_{W}(w)=f_{\varphi(W)}(w)$
		for an unknown injective function $\varphi$. Following identical
		steps to the proof of Theorem 4.1 under HS Assumption 5 we get $\alpha(w)=\varphi^{-1}(w)$,
		and so:
		\begin{align*}
			\tilde{f}_{Y(x_{1},\alpha^{-1}(w))X}(y,x_{2}) & =f_{Y(x_{1},\varphi^{-1}(\alpha^{-1}(w)))X}(y,x_{2})\\
			& =f_{Y(x_{1},w)X}(y,x_{2})
		\end{align*}
		
		Also following the same steps as in Theorem 4.1 we get $Q_{\alpha(\tilde{W})}(\tau)=\phi(Q_{W}(\tau))$
		and $\alpha(\varphi(w))=\phi(w)$. The latter implies $\varphi(w)=\alpha^{-1}(\phi(w))$,
		and so:
		\begin{align*}
			\tilde{f}_{Y(x_{1},\alpha^{-1}(Q_{\alpha(\tilde{W})}(\tau)))X}(y,x_{2}) & =f_{Y(x_{1},\varphi^{-1}(\alpha^{-1}(Q_{\alpha(\tilde{W})}(\tau))))X}(y,x_{2})\\
			& =f_{Y(x_{1},\varphi^{-1}(\varphi(Q_{W}(\tau)))X}(y,x_{2})\\
			& =f_{Y(x_{1},Q_{W}(\tau))X}(y,x_{2})
		\end{align*}
	\end{proof}
	
	\theoremstyle{plain} \newtheorem*{L3}{Lemma A.1} \begin{L3}
		Suppose $X$ is binary, Assumption 10 holds, and for $x=0,1$, $f_{Y(x)|W}$
		is identified up to reorderings of $W$ which can differ for each
		$x$. Define $\bar{s}$ and $\underline{s}$ as in the statement of
		Theorem 6. Then for almost all $w\in\mathcal{W}$: 
		\[
		E[Y(1)-Y(0)|W=w]\in[\underline{s},\bar{s}]
		\]
	\end{L3}

 \begin{proof} By supposition, for each $x\in\mathcal{X}$,
		$f_{Y(x)|W}$ is identified up to reorderings of $W$ which can differ
		for each $x$. More formally, we identify a function $\tilde{f}_{Y(x)|W}$
		so that there is an unknown function $\varphi(w,x)$ so that $\varphi(\cdot,x)$
		is injective with inverse $\varphi^{-1}(\cdot,x)$ for each $x$,
		and $\tilde{f}_{Y(x)|W}(y|w)=f_{Y(x)|\varphi(W,X)}(y|w)$. Note then
		that: 
		\[
		\int_{\mathcal{Y}}y\tilde{f}_{Y(x)|W}(y|w)dy=E[Y(x)|W=\varphi^{-1}(w,x)]
		\]
		The above implies that: 
		\[
		\underset{w\in\mathcal{W}}{\text{ess sup}}\int_{\mathcal{Y}}y\tilde{f}_{Y(x)|W}(y|w)dy=\underset{w\in\mathcal{W}}{\text{ess sup}}E[Y(x)|W=w]
		\]
		And similarly for the essential infima. Substituting into the definitions
		of $\bar{s}$ and $\underline{s}$ we get: 
		\begin{align*}
			\bar{s} & =\underset{w\in\mathcal{W}}{\text{ess sup}}E[Y(1)|W=w]-\underset{w\in\mathcal{W}}{\text{ess sup}}E[Y(0)|W=w]\\
			\underline{s} & =\underset{w\in\mathcal{W}}{\text{ess inf}}E[Y(1)|W=w]-\underset{w\in\mathcal{W}}{\text{ess inf}}E[Y(0)|W=w]
		\end{align*}
		
		Next we will use the above to show that, under rank invariance, $\bar{s}$
		is the supremum of the CATE and $\underline{s}$ is the infimum. Let
		$\{w_{n}\}_{n=1}^{\infty}$ be a sequence in $\mathcal{W}$ so that:
		\[
		E[Y(0)|W=w_{n}]\to\underset{w\in\mathcal{W}}{\text{ess sup}}E[Y(0)|W=w]
		\]
		The above implies that: 
		\[
		P(E[Y(0)|W]\leq E[Y(0)|W=w_{n}])\to1
		\]
		By monotonicity, if $E[Y(0)|W=w]\leq E[Y(0)|W=w_{n}]$ then: 
		\[
		E[Y(1)-Y(0)|W=w]\leq E[Y(1)-Y(0)|W=w_{n}]
		\]
		And so: 
		\[
		P(E[Y(1)-Y(0)|W]\leq E[Y(1)-Y(0)|W=w_{n}])\to1
		\]
		The above then implies: 
		\[
		E[Y(1)-Y(0)|W=w_{n}]\to\underset{w\in\mathcal{W}}{\text{ess sup}}E[Y(1)-Y(0)|W=w]
		\]
		Also by monotonicity, $E[Y(0)|W=w]\leq E[Y(0)|W=w_{n}]$ implies that
		$E[Y(1)|W=w]\leq E[Y(1)|W=w_{n}]$, so by similar reasoning we get:
		\[
		E[Y(1)|W=w_{n}]\to\underset{w\in\mathcal{W}}{\text{ess sup}}E[Y(1)|W=w]
		\]
		Now, using linearity of the expectation we have: 
		\begin{align*}
			\lim_{n\to\infty}\big(E[Y(1)|W=w_{n}]-E[Y(0)|W=w_{n}]\big) & =\lim_{n\to\infty}E[Y(1)-Y(0)|W=w_{n}]\\
			& =\underset{w\in\mathcal{W}}{\text{ess sup}}E[Y(1)-Y(0)|W=w]
		\end{align*}
		By Assumption 10, $\underset{w\in\mathcal{W}}{\text{ess sup}}E[Y(0)|W=w]\leq c<\infty$
		and so we have: 
		\begin{align*}
			& \lim_{n\to\infty}\big(E[Y(1)|W=w_{n}]-E[Y(0)|W=w_{n}]\big)\\
			= & \lim_{n\to\infty}E[Y(1)|W=w_{n}]-\lim_{n\to\infty}E[Y(0)|W=w_{n}]\\
			= & \bar{s}
		\end{align*}
		Combining we get: 
		\[
		\bar{s}=\underset{w\in\mathcal{W}}{\text{ess sup}}E[Y(1)-Y(0)|W=w]
		\]
		Following similar steps we get: 
		\[
		\underline{s}=\underset{w\in\mathcal{W}}{\text{ess inf}}E[Y(1)-Y(0)|W=w]
		\]
		It now follows immediately from the definition of the essential supremum
		and infimum that for almost all $w\in\mathcal{W}$: 
		\[
		E[Y(1)-Y(0)|W=w]\in[\underline{s},\bar{s}]
		\]
	\end{proof}
	
	\begin{proof}[Proof of Theorem A.1] 
		Under conditions i. and ii.
		of Proposition 6 and Assumption 1, 2, and 3, for each $x\in\mathcal{X}$
		we can apply HS Theorem 1 conditional on $X=x$ with $C=Y$. This
		yields identification of $f_{Y|WX}(\cdot|\cdot,x)$, $f_{Z|WX}(\cdot|\cdot,x)$,
		and $f_{W|VX}(\cdot|\cdot,x)$ up to reorderings of $W$ which may
		vary with $x$. Now note that by condition iii. of Proposition 6,
		$f_{Y|WX}(y|w,x)=f_{Y(x)|W}(y|w)$. So $f_{Y(x)|W}$ is identified
		up to reorderings of $W$ which may depend on $x$. We then apply
		Lemma A.1 to get that for almost all $w\in\mathcal{W}$: 
		\[
		E[Y(1)-Y(0)|W=w]\in[\underline{s},\bar{s}]
		\]
		
		For the final result in the theorem note that under conclusion iii.
		of Proposition 6 we have $E[Y(x)|W=w]=E[Y(x)|W=w,X=x]$ for all $x\in\mathcal{X}$,
		and so: 
		\[
		E[Y(1)-Y(0)|W=w]=E[Y(1)-Y(0)|W=w,X=x]
		\]
		Applying the law of iterated expectations: 
		\[
		E[Y(1)-Y(0)|X=x]=E\big[E[Y(1)-Y(0)|W]|X=x\big]
		\]
		We have established that with probability $1$, $E[Y(1)-Y(0)|W,X]\in[\underline{s},\bar{s}]$
		and thus the same holds for the conditional mean of this random variable.	\end{proof}
	
	\begin{proof}[Proof of Theorem A.2] 
		Under conclusions i. ii., and
		iii. of Proposition 7 and Assumptions 2, 7, and 8 we can apply HS
		Theorem 1 within the stratum $x$ of $X$. This yields identification
		of $f_{C|WX}(\cdot|\cdot,x)$, $f_{W|VX}(\cdot|\cdot,x)$, and $f_{Z|WX}(\cdot|\cdot,x)$
		up to reorderings of $W$ which may depend on $x$. Now note that
		by elementary properties of probability densities: 
		\begin{align*}
			& f_{YZ|VX}(y,z|v,x)\\
			= & \int_{\mathcal{W}}f_{Y|WVZX}(y|w,v,z,x)f_{Z|WVX}(z|w,v,x)f_{W|VX}(w|v,x)dw
		\end{align*}
		Proposition 7.iii implies $f_{Y|WVZX}(y|w,v,z,x)=f_{Y|WVX}(y|w,v,x)$.
		In addition, conclusion ii. of Proposition 7 implies $f_{Z|WVX}(z|w,v,x)=f_{Z|WX}(z|w,x)$.
		Substituting we get: 
		\[
		f_{YZ|VX}(y,z|v,x)=\int_{\mathcal{W}}f_{Y|WVX}(y|w,v,x)f_{Z|WX}(z|w,x)f_{W|VX}(w|v,x)dw
		\]
		By Assumption 2 the expression above identifies $f_{Y|WVX}(\cdot|\cdot,\cdot,x)$
		up to reorderings of $W$ (see the steps in the proof of Theorem 4).
		Now note that under conclusion iv. of Proposition 7 we get: 
		\begin{align*}
			f_{Y|WVX}(y|w,v,x) & =f_{Y(x)|WV}(y|w,v)
		\end{align*}
		So $f_{Y(x)|WV}(y|w,v)$ is identified up to reorderings of $W$ which
		may depend on $x$.
		
		We now apply Lemma A.1 within each stratum $v$ of $V$, using Assumption
		11 in place of Assumption 10. We get that for almost all $w\in\mathcal{W}$
		and $v\in\mathcal{V}$: 
		\[
		E[Y(1)-Y(0)|W=w,V=v]\in[\underline{s}(v),\bar{s}(v)]
		\]
		
		Finally, note that under conclusion iv. of Proposition 7 we have that
		for all $x\in\mathcal{X}$ and almost all $w\in\mathcal{W}$ and $v\in\mathcal{V}$
		\[
		E[Y(x)|W=w,V=v]=E[Y(x)|W=w,V=v,X=x]
		\]
		Using the above and applying the law of iterated expectations we get:
		\[
		E[Y(1)-Y(0)|X=x]=E\big[E[Y(1)-Y(0)|W,V]|X=x\big]
		\]
		Since $E[Y(1)-Y(0)|W,V]\leq\bar{s}(V)$ almost surely we have $E\big[E[Y(1)-Y(0)|W,V]|X=x\big]\leq E[\bar{s}(V)|X=x]$
		and similarly for the lower bound. \end{proof}